\documentclass[english,american,superscriptaddress,prd, nofootinbib,11pt,eqsecnum,secnumarabic,aps,floatfix,preprintnumbers]{revtex4-2}
\usepackage{tabularx}
\usepackage{amsmath}  
\usepackage{amsfonts,amssymb,bm}
\usepackage{enumerate}
\usepackage{xcolor}
\usepackage{subfigure}
\usepackage{graphicx} 
\usepackage{xspace}
\usepackage[normalem]{ulem}
\usepackage{braket}
\usepackage{multirow,booktabs}

\usepackage[
margin=0.8in,
]{geometry}
\usepackage[final,pdfencoding=auto]{hyperref} 
\hypersetup{
	colorlinks=true,       
	linkcolor=blue,        
	citecolor=blue,       
	filecolor=magenta,     
	urlcolor=blue         
}

\definecolor{codegreen}{rgb}{0,0.6,0}
\definecolor{codegray}{rgb}{0.5,0.5,0.5}
\definecolor{codepurple}{rgb}{0.58,0,0.82}
\definecolor{backcolour}{rgb}{0.95,0.95,0.92}
\definecolor{nicegreen}{rgb}{0.,0.5,0.}
\definecolor{darkblue}{rgb}{0.0,0,0.5}

\begin{document}
\preprint{MSUHEP-25-019}

\title{Strong Coupling Constant Determination from the new CTEQ-TEA Global QCD Analysis}

\author{Alim Ablat}
\email{alimjanablat@outlook.com}
\affiliation{School of Physics and Electrical Engineering, Kashi University, Kashi, Xinjiang 844000 China\looseness=-1}

\author{Sayipjamal Dulat}
\email{sdulat@hotmail.com}
\affiliation{School of Physics Science and Technology, Xinjiang University, Urumqi, Xinjiang 830046 China\looseness=-1}
\affiliation{Department of Physics and Astronomy, Michigan State University, East Lansing, MI 48824, USA\looseness=-1}

\author{Marco Guzzi }
\email{mguzzi@kennesaw.edu -- Corresponding Author}
\affiliation{Department of Physics, Kennesaw State University, Kennesaw, GA 30144, USA\looseness=-1}

\author{Joey Huston}
\email{huston@msu.edu}
\affiliation{Department of Physics and Astronomy, Michigan State University, East Lansing, MI 48824, USA\looseness=-1}

\author{Kirtimaan Mohan}
\email{kamohan@msu.edu -- Corresponding Author}
\affiliation{Department of Physics and Astronomy, Michigan State University, East Lansing, MI 48824, USA\looseness=-1}

\author{Pavel Nadolsky}
\email{nadolsky@msu.edu -- Corresponding Author}
\affiliation{Department of Physics and Astronomy, Michigan State University, East Lansing, MI 48824, USA\looseness=-1}

\author{Dan Stump}
\email{stump@pa.msu.edu}
\affiliation{Department of Physics and Astronomy, Michigan State University, East Lansing, MI 48824, USA\looseness=-1}

\author{C.-P. Yuan}
\email{yuanch@msu.edu}
\affiliation{Department of Physics and Astronomy, Michigan State University, East Lansing, MI 48824, USA\looseness=-1}

\begin{abstract}
We present a new determination of the strong coupling constant $\alpha_s$ from 
 a global QCD analysis CT25 of parton distribution functions (PDFs) that incorporates high-precision experimental measurements from the Run-2 of the Large Hadron Collider together with a large sample of other measurements over a wide interval of energies. This work addresses two objectives: providing an up-to-date determination of $\alpha_s$ using NNLO calculations and a sensitive nucleon data set within a self-consistent framework, and critically assessing the robustness of the $\alpha_s(M_Z)$ extraction in light of systematic uncertainties as well as correlations of $\alpha_s(M_Z)$ with the functional forms of PDFs and other model parameters. In regard to the uncertainty assessment, we demonstrate that some commonly used criteria, including the dynamical tolerance and Bayesian hierarchical models, may produce significantly different or even unstable estimates for the net $\alpha_s$ uncertainty, and we introduce a concept of \textit{data-clustering safety} that the replicable uncertainty estimates must satisfy. Based on this in-depth examination of the CT25 global hadronic data set using a combination of analysis methods, and after demonstrating a weak correlation between $\alpha_s$ and the functional forms of CT25 PDF parametrizations, we find $\alpha_s(M_Z)=0.1183^{+0.0023}_{-0.0020}$ at the 68\% credibility level.
\end{abstract}

\date{\today}

\maketitle
\flushbottom

\newpage
\tableofcontents

\section{Introduction 
\label{sec:introduction}}

The strong coupling constant $\alpha_s$ is a fundamental parameter of the QCD Lagrangian that enters theoretical computations for all hadronic scattering processes. A precise and accurate determination of $\alpha_s(Q)$ and its dependence on the energy scale $Q$ is of vital importance for the current precision-frontier program at the Large Hadron Collider (LHC), where vast amounts of  data are being gathered and push the boundary of precision measurements to the next level~\cite{dEnterria:2022hzv}. The upcoming high-luminosity phase of the LHC (HL-LHC) will further increase the volume of high-precision data, demanding a matching increase in both the detector capabilities  and precision of theoretical calculations to properly validate relationships between fundamental parameters of the Standard Model (SM), constrain properties of the Higgs boson, and search for new physics.   
For instance, the running of $\alpha_s$ at the highest achieved energies may reveal the presence of new colored particles and influence the scale at which the SM couplings may unify. 
Variations in the observed $\alpha_s(M_Z)$ at the scales of order of the $Z$ boson mass, $M_Z$, can affect projections for the range of energies at which the electroweak (EW) vacuum might shift from a metastable to a stable configuration.

Hence, the QCD coupling and its estimated uncertainty play a pivotal dual role of characterizing the universal force of nature and, at the same time, serving as a precise theoretical input for many standard candle processes in ongoing experiments. In particular, top-quark pair and Higgs boson production cross sections are very sensitive to the assumed $\alpha_s(M_Z)$. The associated uncertainty, which is correlated in hadron-collider predictions with the PDF uncertainty, can potentially exceed those due to the scale dependence and experimental errors~\cite{Salam:2017qdl,Dulat:2018rbf}.  

The Particle Data Group (PDG) currently reports the uncertainty on $\alpha_s(M_Z)$ of about 0.8\%, resulting in a value of $\alpha_s(M_Z) = 0.118 \pm 0.0009$ ~\cite{rpp_2024_PhysRevD.110.030001}. This result is derived from a compilation of lattice and non-lattice results, which independently provide compatible central values of $\alpha_s(M_Z)$ and of its uncertainty. The corresponding relative uncertainty exceeds that achieved for the fundamental couplings of electroweak and gravitational interactions. The new twist is that the latest determinations of $\alpha_s$ from high-statistics experimental measurements and lattice QCD may individually quote a comparable and --- perhaps in the future --- a smaller uncertainty than the PDG world-average uncertainty of $\pm 0.0009$. Along these lines, the ATLAS collaboration measured $\alpha_s(M_Z) = 0.1183\pm0.0009$~\cite{ATLAS:2023lhg} using data of $Z$ boson transverse-momentum distributions at $\sqrt{s}=8$ TeV. The CMS collaboration simultaneously extracted $\alpha_s(M_Z)$ and PDFs~\cite{CMS:2024trs} from a global QCD analysis at NNLO using double-differential cross section measurements of inclusive jet production at $\sqrt{s}=$ 2.76, 7, 8, and 13 TeV combined with deep inelastic scattering (DIS) data from HERA. They obtained $\alpha_s(M_Z) = 0.1176^{+0.0014}_{-0.0016}$. Other recent relevant $\alpha_s$ determinations from the CMS and ATLAS collaborations can be found in Refs.~\cite{CMS:2024hwr,CMS:2024mlf,ATLAS:2023tgo,ATLAS:2023lhg}. Several determinations from deep inelastic scattering (DIS) at $ep$ collider HERA find a somewhat lower $\alpha_s$ value than at the LHC, such as $\alpha_s(M_Z)=0.1142\pm 0.0028$ determined by H1 collaboration in a simultaneous fit of $\alpha_s$ and proton PDFs \cite{H1:2017bml}. Independently, lattice QCD determines $\alpha_s$ from several types of nonperturbative computations. The Flavor Lattice Averaging Group (FLAG) systematizes and cross-validates these determinations. It recently reports $\alpha_s(M_Z)=0.1183\pm 0.0007$ \cite{FLAG:2024oxs}.

The studies of $\alpha_s$ thus enter a new phase when its precision may eventually be pushed to 0.1\% across diverse observations in the next two decades \cite{dEnterria:2022hzv}, while in the meantime novel challenges will need to be confronted in both the individual $\alpha_s$ measurements and their combination. It is expected that, at this precision level, the growing complexity of involved systematic factors raises the chance for misestimations. When the misestimations go unnoticed -- an unwanted but common occurrence encountered in complex studies across many disciplines \cite{ReproducibilityReplicabilityNAS:2019} -- the uncertainty tends to be underestimated; some measurements of the parameter are not replicated by its independent determinations. 
When replicability degrades, a PDG or another combination  
may need to inflate the stated uncertainties of measurements in order to make them statistically compatible. As an illustration, the Committee on Data of the International Science Council inflates the errors of highly precise measurements of Newton's gravitational constant $G$ by a factor of 3.9 in order to meaningfully combine them \cite{Codata2022GravConstant:2025}. Within its small span, the combined uncertainty band on $G$ does not overlap significantly with those of several most precise measurements, meaning it does not replicate them. This indicates a general trend that the measurements of fundamental constants eventually experience the tradeoff between the precision and replicability, when raising the former depletes the confidence in the latter. How far can the precision measurements of $\alpha_s$ go while remaining replicable? What can be done to control the uncertainty estimates and postpone the onset of non-replicability?

This article will critically explore these questions in the context of the precision determination of the $\alpha_s(M_Z)$ from the new CTEQ Tung Et Al. (CTEQ-TEA) global QCD analysis of nucleon scattering data, named ``CT25''~\cite{ct25:2025xxx}. The primary goal of the global QCD analysis is to determine parametrizations of parton distribution functions (PDFs), as well as $\alpha_s$ and other QCD parameters, from a simultaneous fit of a large, diverse collection of experimental measurements. The methodology of the global analysis offers the most complete view of the  {\it correlated} constraints on $\alpha_s$, heavy-quark masses, and the PDFs arising from a variety of scattering experiments with different systematic errors, with {\it simultaneous} dependence on all of them. This completeness gives the global analysis an edge over the standalone $\alpha_s$ determinations from the nucleon data, which typically neglect at least some of the correlation with the PDFs or other experiments. As such, and given the growing dominance of hadron scattering processes in experimental $\alpha_s$ determinations, the global analysis of PDFs is well suited for tests of uncertainty quantification and replicability in the $\alpha_s$ studies. Many combined analyses of the PDFs and $\alpha_s$ have been performed by our CTEQ-TEA group \cite{Lai:2010nw,Hou:2019efy} and other groups \cite{Alekhin:2017kpj,Gizhko:2017fiu,Ball:2018iqk,Cridge:2021qfd,Cridge:2023ztj,Alekhin:2024bhs} over the years.
The strong coupling $\alpha_s$ has recently been extracted from global QCD analyses, employing an approximate PDF evolution at the next-to-next-to-next-to-leading order (aN$^3$LO)~\cite{Cridge:2024exf} in QCD by the MSHT group, and from an analysis using an aN$^3$LO QCD evolution combined with the photon PDF at the next-to-leading order (NLO) in QED~\cite{Ball:2025xgq} by the NNPDF collaboration. 

The CT25 global analysis determines $\alpha_s$  at the next-to-next-to-leading order (NNLO) in QCD by employing an improved methodology as well as an extended global data set in which precision LHC Run-2 measurements with high integrated luminosity augment those included in the CT18 study~\cite{Hou:2019efy}. The novel feature of our study is that we determine the central value and uncertainty of $\alpha_s$ by several methods to obviate the dependence on the correlation with PDFs, selection of experiments, and systematic uncertainties in the experiments. 

We do  not resort to a single prescribed definition of the global PDF uncertainty (tolerance \cite{Kovarik:2019xvh}) to estimate the allowed $\alpha_s$ range, in contrast to the previous combined PDF+$\alpha_s$ fits. Instead, we determine the span of the $\alpha_s$ values consistent with the global data set at NNLO in several ways:
(i) by generating the $\chi^{2}(\alpha_{s})$ profile from a single fit, (ii) by comparing the $\alpha_s$ values from many fits with independent PDF parametrizations, and (iii) by combining the constraints from individual experiments according to several criteria to assess the uncertainty. The final $\alpha_s$ result is obtained by combining several such extractions. 

The rest of this manuscript is organized as follows. 
In Sec.~\ref{sec:pdfs}, we briefly describe the methodology and advancements in the CT25 global PDF analysis. We summarize the new data sets included in the CT25 baseline, the definition of the figure-of-merit function $\chi^2$, and the implementation of correlated systematic errors provided by the experiments.
Section~\ref{sec:results-P1} explores the impact of the choice of PDF parametrizations on the $\alpha_s$ determination. We also review how $\alpha_s(M_Z)$ can be determined from the $\chi^2$ scan over $\alpha_s(M_Z)$, and which criteria are applicable to determine the $\alpha_s(M_Z)$ error with this methodology. 

Sec.~\ref{sec:results-P2} critically examines the impact of the implementation of experimental systematic uncertainties on the extracted value of $\alpha_{s}(M_{Z})$, and it discusses the handling of some observed disagreements among data sets in the baseline.
Section~\ref{sec:Tolerances} is dedicated to the exploration of uncertainty prescriptions (tolerances) according to the commonly used global and dynamical implementations. For the latter definition, we examine how the user-specified clustering of published data sets modifies the quantitative error estimates.
We point out that either the dynamical tolerance in the Hessian method or $K$-folding in the Monte-Carlo method may produce uncertainties that depend on the user-chosen clustering of data sets. We introduce the concept of data clustering safety that must be addressed in such situations. 

Sec.~\ref{sec:BayesianModels} describes Bayesian statistical models utilized for the combination of the results for $\alpha_s(M_Z)$ and its uncertainties, while Sec.~\ref{sec:combination} obtains and compares the uncertainties using a variety of methods. This Section quotes our final $\alpha_s$ estimate. Sec.~\ref{sec:conclusion} contains our conclusions. 
Technical details of several discussions are relegated to five Appendices. The default scale $M_Z$ is to be assumed when the scale of $\alpha_s$ is omitted. 


\section{Relevant aspects of the CT25 global analysis
\label{sec:pdfs}}

\subsection{Selection of data sets \label{sec:DataSets}}
As in many general-purpose CTEQ-TEA fits, the CT25 NNLO global analysis has taken place over several years and now culminates with the release of the official CT25 PDF ensembles that update the previous generation of the PDFs \cite{Hou:2019efy} consisting of the default CT18 NNLO error set and three alternative, yet equally acceptable PDF ensembles, CT18Z, A, and X.  A significant part of this effort went into an exploration of a large number of new candidate measurements for the extension of the CT18 baseline data set, focusing primarily on high-precision production of lepton pairs, hadronic jets, and top quark-antiquark pairs in the LHC Runs 1 and 2. We have documented these preliminary explorations in three publications \cite{Sitiwaldi:2023jjp, Ablat:2023tiy, Ablat:2024uvg}. Based on these studies, we selected a number of precise and mutually consistent data sets that are included in the final CT25 release.
Several data sets included in CT18 have been updated or removed in the final CT25 fit, for the reasons that will be explained in the text.

The newly added LHC data sets in CT25 are listed in the upper part of Table~\ref{Tab:LHCnewData} and reviewed in Appendix~\ref{app:NewData}. As its data baseline, CT25 started with the CT18A~\cite{Hou:2019efy} global data set, which already included a combination of measurements in deep-inelastic scattering and production of lepton pairs, hadronic jets, and top-quark pairs at both fixed-target and collider energies. Notably, the CT18A baseline included a precision data set on $W$ and $Z$ production in the ATLAS experiment at 7 TeV \cite{Aaboud:2016btc}, which offered the first evidence for the enhanced strangeness PDF at $x\approx 0.02$, in some disagreement with charged-current DIS experiments. More recent LHC measurements included in the CT25 fit are consistent with this strangeness enhancement, albeit with somewhat reduced magnitude.

The inclusion of the new LHC measurements, added on top of the CT18A baseline {\it without replacing or discarding any of the previously fitted experiments}, increased the number of data points, $N_\textrm{pt}$, by 769, the majority of which were from inclusive jet production and Drell-Yan triple-differential $Z$-boson production at ATLAS, cf. Table~\ref{Tab:LHCnewData}. This version of the CT25 fit, which we will call ``CT25prel'', has the largest $N_\textrm{pt}=4450$ among the various CT25 candidate fits. This data set is not unique, however: other well-motivated data combinations are possible, as was also  the case for CT18. When all data sets are retained in the CT25 baseline, we observe a reasonable, although not ideal, agreement among the global data, with significant pulls on the PDFs from the newly-fitted data. The most precise new experiments in the CT25 fits tend to exhibit an increase in $\chi^2/N_\textrm{pt}$ in a pattern similar to that observed in the CT18 global analysis. The new Drell-Yan and $t\bar t$ production data consistently prefer a softer gluon at intermediate-to-large $x$ ($0.05\leq x\leq 0.7$), while inclusive jet production data pull the gluon in the opposite direction in the large-$x$ range. Overall, the cumulative effect of these measurements, when they are combined, amount to having a softer gluon at  intermediate-to-large $x$, as compared to CT18A.

To investigate the range of valid PDF solutions in this new setting, we explored other related, also acceptable, branches of the CT25prel fit. The alternative choices made in these fits reflect the complexities present within the global fitting framework, many of which were highlighted in the preceding CT18 publication \cite{Hou:2019efy}, as well as in a review \cite{Kovarik:2019xvh}.\footnote{In the context of CT18, we opted to publish four PDF ensembles to quantify some of these possibilities.} 
In the end, we obtained the default CT25 fit, or just ``CT25'', as follows.
Compared to CT18A and CT25prel, we replaced the H1 measurements of charm and bottom semi-inclusive structure functions \cite{Abramowicz:1900rp,Aktas:2004az} by their HERA combination \cite{H1:2018flt}, which moderately increased the corresponding $\chi^2$ without substantially modifying the PDFs. We added the E906/SeaQuest measurement of the $\sigma_{pd}/(2\sigma_{pp})$ ratio in fixed-target Drell-Yan pair production \cite{SeaQuest:2021zxb}. We removed four data sets (336 points) on inclusive charged-current DIS on an iron target from the CDHSW \cite{Berge:1989hr} and CCFR \cite{Yang:2000ju,Seligman:1997mc} collaborations, as they are in some tension with each other and increase dependence of CT25 results on the nuclear modifications of proton PDFs.\footnote{The default CT25 fit nevertheless retains CCFR and NuTeV data sets on dimuon production in CC DIS on the isoscalar target, as those provide unique constraints on the strangeness PDF at $x>0.05$. In both CT18 and CT25 fits, we apply a correction from \cite{Rondio:1993mf} to approximate nuclear modifications in such experiments.} The weakly sensitive determination of the longitudinal structure function $F_L(x_B,Q^2)$ by the H1 collaboration~\cite{H1:2010fzx} was also omitted without a noticeable impact on the PDFs. Finally, we replaced the CMS jet production datasets at 7, 8, and 13 TeV by their updated versions~\cite{CMS:2024trs} that incorporate improved evaluations of systematic uncertainties and include statistical correlation matrices.

\begin{table}[p]
\caption{The upper part of the table lists the newly added data sets in the CT25 analysis, contributed by measurements of lepton-pair, top-quark-pair, and single-inclusive jet production from the LHC Runs 1 and 2. 
For these experiments, we report the $\chi^2/N_\textrm{pt}$ for the  default CT25 and prior CT18A PDF ensembles, assuming $\alpha_s(M_Z)=0.118$. Section~\ref{sec:chisquare} provides the definition of $\chi^2$. The errors are asymmetric PDF uncertainties computed using a master formula in Ref.~\cite{Lai:2010vv} and the error PDFs defined at 68\% CL, according to the two-tier PDF tolerance employed in CT18.
The lower part of the table shows the total numbers of points, $\chi^2/N_{\textrm{pt}}$, and approximate preferred $\alpha_s$ in the default and preliminary CT25 fits described in the text, as well as in the CT18A fit.} 
	\centering
	\begin{tabular}{clccc}
		\hline\hline
        \multicolumn{5}{c}{New and updated LHC data sets in the CT25 NNLO analysis} \\
        \hline
		ID & Experiment & $N_{\rm pt}$ &  $\chi^{2}/N_{\rm pt}$ for $\alpha_s(M_Z)=0.118$, CT25 (CT18A) & \\
		\hline            
		\multicolumn{5}{c}{Lepton pair production} \\
		\hline
		211  & ATLAS 8 TeV $W$~\cite{ATLAS:2019fgb}       & 22  & $	  2.04^{	+0.65	}_{-0.52} $ ( $ 4.35^{	+2.65	}_{ -2.37} $)  \\
        212  & CMS 13 TeV $Z$~\cite{CMS:2019raw}          & 12  & $	  2.02^{	+1.03	}_{-1.23} $ ( $ 2.12^{	+3.77	}_{ -0.17} $)  \\
        214  & ATLAS 8 TeV $Z$ 3D~\cite{ATLAS:2017rue}    & 188 & $	  1.16^{	+0.14	}_{-0.06} $ ( $ 1.22^{	+0.34	}_{ -0.14} $)  \\
		215  & ATLAS 5.02 TeV $W,Z$~\cite{ATLAS:2018pyl} & 27  & $	  0.62^{	+0.10	}_{-0.09} $ ( $ 0.77^{	+0.34	}_{ -0.09} $)  \\
		217  & LHCb 8 TeV $W$~\cite{LHCb:2016zpq}         & 14  & $	  1.39^{	+0.34	}_{-0.23} $ ( $ 1.52^{	+0.49	}_{ -0.36} $)  \\
		218  & LHCb 13 TeV $Z$~\cite{LHCb:2021huf}        & 16  & $	  1.09^{	+0.49	}_{-0.39} $ ( $ 1.24^{	+1.03	}_{ -0.38} $)  \\
		\hline
	\multicolumn{5}{c}{Inclusive jet production} \\
		\hline
        553 & ATLAS 8 TeV jets~\cite{ATLAS:2017kux}   &171 & $1.60^{	+0.10	}_{-0.06}$  (  $1.57^{	+0.12	}_{ -0.07}$ )   \\
		554 & ATLAS 13 TeV jets~\cite{ATLAS:2017ble}  &177 & $1.32^{	+0.06	}_{-0.05}$  (  $1.26^{	+0.09	}_{ -0.03}$ )   \\
555 & CMS   13 TeV jets~\cite{CMS:2021yzl,CMS:2024trs}    &78  & $1.13^{	+0.11	}_{-0.04}$  (  $1.29^{	+0.23	}_{ -0.17}$ )  \\
556& CMS 7 TeV jets~\cite{CMS:2012ftr,CMS:2014nvq,CMS:2014qtp,CMS:2024trs}  & 118  & $0.74^{	+0.02	}_{-0.03}$  (  $0.7^{	+0.12	}_{ -0.03}$ ) \\
557 & CMS   8 TeV jets~\cite{CMS:2016lna,CMS:2024trs}  & 164 & $1.17^{	+0.12	}_{-0.06}$  (  $1.16^{	+0.18	}_{ -0.06}$ )  \\
		\multicolumn{5}{c}{ $t\bar{t}$ production at 13 TeV} \\
		\hline
		521 & ATLAS all-hadronic $y_{t\bar{t}}$ ~\cite{ATLAS:2020ccu}                                           & 12 &  $1.11^{	+0.07	}_{-0.07}$ ($ 1.06^{	+0.10	}_{ -0.07} $ )   \\
		528 & CMS dilepton  $y_{t\bar{t}}$~\cite{CMS:2018adi}                                                   & 10 &  $1.28^{	+0.43	}_{-0.44}$ ($ 1.04^{	+0.78	}_{ -0.44} $ )   \\
		581 & CMS lepton+jet  $m_{t\bar{t}}$~\cite{CMS:2021vhb}                                                & 15 &  $1.13^{	+0.33	}_{-0.31}$ ($ 1.36^{	+0.89	}_{ -0.48} $ )   \\
        587 & ATLAS lepton+jet  $m_{t\bar{t}}+y_{t\bar{t}}+y^B_{t\bar{t}}+H_T^{t\bar{t}}$ ~\cite{ATLAS:2019hxz}& 34 &  $1.07^{	+0.19	}_{-0.13}$ ($ 0.94^{	+0.24	}_{ -0.10}$) \\

      \hline \hline 
       \end{tabular}
       \quad\\ \vspace{1\baselineskip}
      	\begin{tabular}{clccc}
       \hline \hline 
		\multicolumn{5}{c}{All experiments} \\
      \hline
		& PDF ensemble & $N_{\rm pt}$ &  $\chi^{2}/N_{\rm pt}$ for $\alpha_s(M_Z)=0.118$& Preferred $\alpha_s(M_Z)$ \\
      \hline      
      & CT25 NNLO, with default data set~\cite{ct25:2025xxx}& 4066 
      &  $1.20   \pm      0.009$  & $\approx 0.1185 $ \\
      & CT25prel, with augmented CT18A data set& 4450 
      & $1.20   \pm      0.010$  & $\approx 0.1177$\\
        & CT18A NNLO ~\cite{Hou:2019efy}& 3674 & 
        $1.19  \pm       0.013$ & $\approx 0.1165 $\\
   \hline\hline
	\end{tabular}
\label{Tab:LHCnewData}
\end{table}

Turning back to Table~\ref{Tab:LHCnewData}, the last column in its upper portion reports the $\chi^2/N_{\rm pt}$ values for each  experiment using the PDFs from the default CT25 NNLO and from the CT18A NNLO global fits (with the latter in parentheses). The errors express the corresponding asymmetric PDF uncertainties that are evaluated at the 68\% credibility level (CL).\footnote{The PDF fits base themselves on the Bayesian paradigm, and hence the interpretation of their errors on the PDFs and $\alpha_s$ as ``credibility intervals'' is appropriate. That said, these errors are also used to predict the frequency of future measurements using the determined parameters, in which case the users may view them as ``confidence intervals''.} Fitting the new experiments in the CT25 framework generally improves their $\chi^2$.

The lower portion of Table~\ref{Tab:LHCnewData} compares the total numbers of points and $\chi^2/N_\textrm{pt}$ for the default CT25, CT25prel, and their CT18A predecessor. The last column of the lower part also shows the {\it approximate} preferred value of $\alpha_s$ in these three fits, which is further adjusted by modifications discussed in the rest of the article. 

Upon the addition of the LHC data sets in CT25prel, the reference $\alpha_s$ increased from a range 0.1164-0.1169 in the CT18 series to $\approx 0.118$. To a large extent, this change is due to the preference for a softer gluon at the intermediate-to-large $x$, which is compensated by some increase in $\alpha_s$. 

In the default CT25 fit, the reference $\alpha_s$ is further increased to $\approx 0.1185$ primarily because of the removal of the CDHSW inclusive charged-current DIS data set \cite{Berge:1989hr} on iron. The exclusion of the counterpart CCFR measurement had no appreciable effect, as it is known to be largely consistent with $\alpha_s(M_Z)\approx 0.118$\cite{Alekhin:1998df,Kataev:1999bp,Kataev:2001kk,Kataev:2001fv}. In the CT18 publication, we emphasized that the CDHSW data set is generally fitted well, yet its $Q$ dependence differs from that of an analogous CCFR observation \cite{Seligman:1997,CCFRNuTeV:1995oun}. 

The Lagrange Multiplier scans in the context of CT18 (see its Fig.~21 in Ref.~\cite{Hou:2019efy}) indicated that CDHSW prefers a far larger gluon $g(x,Q)$ at $x>0.1$ than the other experiments. This preference, in turn, suppresses the preferred $\alpha_s$. For this reason, the two CDHSW data sets were included in CT18/CT18A/CT18X, but not in CT18Z. Aside from the impact on the fitted value of $\alpha_s$, the removal of the inclusive neutrino-iron DIS data sets from the CT25prel PDF fit slightly (well within PDF uncertainties) reduced $g$ and increased $u$ and $d$ PDFs at $x>0.1$. Overall, these CDHSW and CCFR experiments do not manifestly disagree with the rest of the data.

Our general observation is then that the described revisions in the CT25 data set, as compared to the CT18A data set, tend to elevate the preferred $\alpha_s$ to above 0.118 {\it for the default settings of the global fit.} We further emphasize that this trend may change considerably under reasonable modifications in the fit's settings, reflecting in part {\it inconsistencies (tensions) among the input data sets}, and in part the {\it modeling of experimental and theoretical systematic uncertainties}. 

Many CPU-intensive studies were carried out with the CT25prel data set, and were not repeated for the (slightly different) CT25 data set. In the rest of this article, we will present studies with either CT25prel or with CT25, without loss of generality.

\subsection{The CTEQ-TEA goodness-of-fit function}
\label{sec:chisquare}

In the exploration of uncertainties, we will repeatedly turn to the goodness-of-fit (GOF) function $\chi^2 \equiv -2 \ln P(T|D)$, minimized with respect to all free parameters in the fit to maximize the posterior probability $P(T|D)$. The CT25 analysis implements it as
\begin{equation}
\chi^2 = \chi^2_R + \chi^2_\textrm{LM},
\label{chi2total}
\end{equation}
where $\chi^2_R$ quantifies the agreement of theory with experiments, and $\chi^2_\textrm{LM}$ is a generally small contribution imposing theoretical priors on the PDFs. The latter is constructed from Lagrange-Multiplier (LM) contributions whose role is to (i) enforce physical behavior of PDF flavor ratios in the extrapolation regions with poor constraints from the data and (ii) enable a ``ridge'' (L2) regularization on a few poorly constrained, yet correlated free parameters. 

In a fit with  $N_E$ data sets, the first term in Eq.~(\ref{chi2total}), 
\begin{equation}
\chi^2_R \equiv \sum_{E=1}^{N_E} \chi_E^2, 
\label{chi2R}
\end{equation}
is a sum of contributions $\chi^2_E$ from individual data sets $E$ given by~\cite{Hou:2019efy,Lai:2010vv}
\begin{align}
& \chi^2_{E} (\bm{a},\bm{\lambda}, \alpha_s, \{m_q\}) = \nonumber \\
& \sum_{k=1}^{N_\textrm{pt}^{(E)}} \frac{1}{s_k^2}\left(D_k -T_k\left(\bm{a}, \alpha_s, \{m_q\}\right) -\sum_{\alpha=1}^{N_\lambda^{(E)}} \beta_{k\alpha}\left(\bm{a},\alpha_s, {m_q}\right) \lambda_\alpha  \right)^2 + \sum_{\alpha=1}^{N_\lambda^{(E)}} \lambda_\alpha^2\,. 
\label{chi2E-def}
\end{align}
Here $D_k$ and $T_k$ are the $k$-th central experimental and theory values, respectively, $s_k =\sqrt{s^2_{k,\textrm{stat}}+s^2_{k,\textrm{uncor sys}}}$ is the total uncorrelated error on the measurement $D_k$, and $\beta_{k\alpha}^{(E)}$ is the correlation matrix of systematic uncertainties. 
$\bm{a} \equiv\{a_1,a_2\dots,a_{N_\textrm{pdf}}\}$ and $\bm{\lambda}\equiv\{\lambda_1, ...\lambda_{N_\lambda^{(E)}}\}$ are the vectors of PDF parameters and systematic nuisance parameters of $E$, with the latter commonly assumed to be random and distributed according to the standard normal distribution $\mathcal{N}(0,1)$. Equation~(\ref{chi2E-def}) indicates that $\chi^2$ generally depends on QCD parameters -- $\alpha_s$ and quark masses $m_q$ -- both through the theoretical predictions $T$ and the experimental correlation matrices $\beta_{k\alpha}$, although this dependence of the latter is usually neglected. 


In our study, we minimize the total $\chi^2$ with respect to $\alpha_s$ and $\bm{a}$, while using constant $\{m_q\}$ and profiling \cite{Pumplin:2002vw} the nuisance parameters $\bm{\lambda}$ for every $\{\alpha_s, \bm{a}\}$ combination. We will find it helpful that, at its global minimum for each $\alpha_s$, the $\chi^2$ can be expressed in terms of the best-fit PDF parameter combination $\bm{a}_0$ as (see Eq. (B8) in \cite{Hou:2019efy})
\begin{equation}
    \min_{\bm{a}} \chi^2 (\bm{a}, \alpha_s, \{m_q\}) \equiv \chi^2 (\bm{a}_0, \alpha_s, \{m_q\}) = \sum_E \left[ 
    \underbrace{\sum_{k=1}^{N_\textrm{pt}^{(E)}} [r_{0,k}^{(E)}]^2}_{\equiv D^2} + \underbrace{\sum_{\alpha=1}^{N_\lambda^{(E)}} [\lambda^{(E)}_{0,\alpha}]^2}_{\equiv R^2}\right],
    \label{minchi2}
\end{equation}
i.e., it is a quadrature sum of the best-fit data residuals $r_{0,k}^{(E)}$ and best-fit nuisance parameters for each experiment $E$. 
Their explicit formulas are \cite{Hou:2019efy}
\begin{align}
& r^{(E)}_{0,k} = \frac{1}{s_k}\left( D_k - T_k(\bm{a}_0) -\sum_{\alpha=1}^{N_\lambda^{(E)}} \beta_{k\alpha} \lambda^{(E)}_{0, \alpha} \right), \mbox{ and} \\
& \lambda^{(E)}_{0,\alpha} = \frac{1}{s_k^2}\left( D_k - T_k(\bm{a}_0) \right) \beta_{k\gamma}\mathcal{A}^{-1}_{\gamma\alpha}, \mbox{ with } 
\mathcal{A}_{\alpha_1 \alpha_2 }=\delta_{\alpha_1\alpha_2}+\sum_{k=1}^{N_\textrm{pt}^{(E)}}\frac{\beta_{k\alpha_1}\beta_{k\alpha_2}}{s_k^2},
\label{res-def}
\end{align}
and all terms on the right-hand side corresponding to experiment $E$.
Furthermore, since $r_{0,k}^{(E)}$ and $\lambda^{(E)}_{0,\alpha}$ are correlated only through a relatively small number of PDF parameters, especially across separate experiments, in the first approximation they all can be assumed to be independently and normally distributed, $r_{0,k}^{(E)} \sim \mathcal{N}(0,1)$ and $\lambda^{(E)}_{0,\alpha} \sim \mathcal{N}(0,1)$. 
This independence opens the possibility of reclustering the individual data points in the total $\chi^2$ in Eq.~(\ref{minchi2}), which will be pursued in Sec.~\ref{sec:Tolerances}. Furthermore, in Sec.~\ref{sec:reclustering} we can examine separate reclustering of sums of all best-fit residuals and nuisance parameters, denoted as $D^2$ and $R^2$, respectively.

This reclustering procedure does not affect the best fit but can be useful for understanding stochastic effects on the resultant uncertainty, especially in connection with a dynamic tolerance prescription.

\subsection{Limitations of existing models for correlated systematics\label{sec:CorrelationLimitations}}
Several aspects of the implementation of correlated systematic errors through $\beta_{k\alpha}$ in our $\chi^2$ introduced above will be relevant for interpreting the results that will follow.  
\begin{enumerate}
\item We first note that the total number $N_\lambda$ of experimental nuisance parameters has grown from at most a few tens in the early global fits to more than 2000 in the CT25 analysis. That is, $N_\lambda$ is now close to a half of the total data points $N_\textrm{pt}$ due to the tendency of the latest experiments to provide large correlation or covariance matrices. In the best fit, many of these nuisance parameters are not normally distributed, but instead cluster close to zero, leading to a spike in the distribution.
\item Each nuisance parameter is included with a prior constraint, essentially an extra datapoint preventing the parameter's completely arbitrary variations.  
Considerable attention has been paid recently to the fact that, by using the linear Gaussian models to publish their correlated systematic uncertainties, experiments unavoidably leave out critical information or introduce spurious assumptions that may proliferate when the number of priors is large. There are a couple such points, first, related to the standard assertion that all systematic nuisance parameters follow the standard normal distribution, $\lambda^{(E)}_{\alpha} \sim \mathcal{N}(0,1)$ in the $\chi^2$ definition in Eq.~(\ref{chi2E-def}). Not only is it highly unlikely that all $\lambda_\alpha$ parameters will be thus distributed, but there are also known important cases where this assumption is false, e.g., for so-called two-point uncertainties that take the difference of at most two estimates, for instance, provided by two parton shower programs, as a measure of a possible uncertainty \cite{Bailey:2019yze}.  
\item The second point of the information loss emerges from the common practice of publishing only {\it normalized} correlation matrices, $\sigma_{k\alpha}$, from which the actual correlation matrices, 
$\beta_{k\alpha}=X_k \sigma_{k \alpha}$ in Eq.~(\ref{chi2E-def}), must be reconstructed by multiplying by unspecified reference values $X_k$ that are chosen within the PDF fit itself. The $X_k$ values are of order of either the central data values, $D_k$, or respective theoretical values, $T_k$. They do not need to coincide with either; the prescription for choosing $X_k$ is a delicate part of the fitting model \cite{Gao:2013wwa,Ball:2012wy}, approached differently by each PDF fitting group \cite{Ball:2009qv,Gao:2013xoa}, who try to reasonably fill the lacunas in the correlation models they get.
\end{enumerate}

How do these considerations impact the precise determination of the QCD coupling constant? The next section offers some answers, after discussing another potential source of uncertainty due to the freedom in the PDF parametrization.


\section{Sources of uncertainties on QCD coupling
\label{sec:SourcesOfUncertainties}
}

\subsection{The PDF parametrization uncertainty is small...
\label{sec:results-P1}}

It has been pointed out over years, e.g., in ~\cite{Lai:2010nw,Hou:2019efy,Forte:2020pyp,Forte:2025pvf}, that 
QCD parameters determined from a global fit are mutually dependent on the PDFs and thus must be fitted simultaneously with the PDFs. Within the global fit, the correlation between the $\alpha_s$ and PDFs can be captured either by fitting them together or by scanning the $\chi^2(\alpha_s, \mathbf{a}_0(\alpha_s))$ over $\alpha_s$, while minimizing the $\chi^2$ with respect to the PDFs for each $\alpha_s$ value. Either approach can be used to determine the best-fit combination of $\alpha_s$ and PDFs together with their uncertainties. We will employ both, especially the latter one -- the $\alpha_s$ scan.

Perhaps unexpectedly, our study found a low correlation between the best-fit $\alpha_s$ and the functional forms of PDF parametrizations in the vicinity of the overall global $\chi^2$ minimum. The consequence is that the choice of the PDF parametrizations has a low impact on the $\alpha_s$ and its error, among the candidate functional forms of the CT25 fit that render the lowest achieved values of $\chi^2$. The variation of the best-fit $\alpha_s$ caused by the choice of the parametrization forms is comparable to the uncertainty corresponding to a $\Delta \chi^2=1$ change in the global $\chi^2$. 
We will see that this is quite a small uncertainty compared to other sources. 

\begin{figure}[tb]
    \centering
    \includegraphics[width=0.75\linewidth]{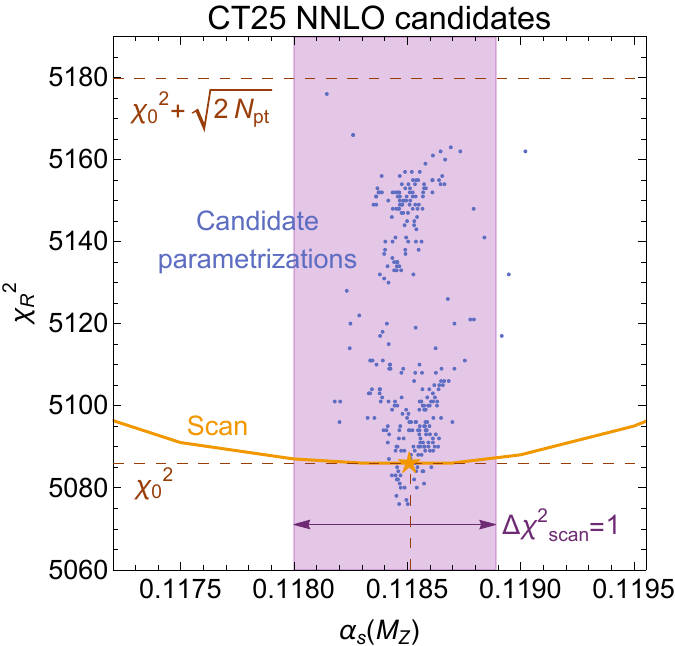}
    \caption{
    $\chi^2$ of the data versus $\alpha_s(M_Z)$ from the scan of the nominal CT25 PDF functional forms and PDF+$\alpha_s$ fits with 287 alternative parametrizations. 
    \label{fig:alphas_scan_290forms}}
\end{figure}

Figure \ref{fig:alphas_scan_290forms} illustrates the low correlation of best-fit $\alpha_s$ and PDFs by comparing the $\chi^2$ dependence from an $\alpha_s$ scan for the functional forms of the nominal CT25 fit and 287 fits with alternative PDF parametrizations explored as a part of the full CT25 exercise. In this plot, we focus on the component of the total $\chi^2$ consisting from experimental log-likelihoods and denoted by $\chi^2_R$ as in Eq.~(\ref{chi2total}), and neglecting the prior component $\chi^2_\textrm{LM}$ that is relatively constant in this comparison. All fits in this figure are done with identical settings that are close both to the CT25prel and final CT25 ones introduced in Sec.~\ref{sec:DataSets}. Namely, the shown fit with the nominal CT25 parametrization achieves $\min\chi^2_R\equiv \chi^2_0=5087$ for $N_\textrm{pt}=4260$ data points. Compared to the CT25prel fit, it excludes the CDHSW heavy-nucleus CC DIS data set \cite{Berge:1989hr}, which prefers a low $\alpha_s$, as well as the weakly sensitive H1 $F_L$ data set~\cite{H1:2010fzx}. It retains other experiments of the CT25prel baseline. 

The orange star in Fig.~\ref{fig:alphas_scan_290forms} indicates the best-fit combination $\{\alpha_s\approx 0.1185, \chi^2_R=5087\}$ obtained with the nominal CT25 parametrization. The orange curve indicates the $\chi^2_R$ dependence from the $\alpha_s$ scan with this parametrization. This curve closely follows the parabolic behavior. The vertical magenta band indicates the interval of $\alpha_s$ values corresponding to the increase $\Delta \chi^2_R \leq 1$ from the $\alpha_s$ scan, measured with respect to the minimal $\chi^2_R$ on this quasi-parabola. 

The blue dots indicate the best-fit $\{\alpha_s, \chi^2_R\}$ combinations from simultaneous PDF+$\alpha_s$ fits that use 287 variants of PDF parametrization forms. In these parametrizations, both the functional forms and the number of parameters are varied. A small number of candidate fits in Fig.~\ref{fig:alphas_scan_290forms} have a lower $\chi^2_R$ (slightly better agreement with data) than the nominal CT25 parameterization (the orange star). Yet they have a slightly worse overall $\chi^2$ due to the additional LM penalties in the extrapolation regions, e.g. because they result in less physical asymptotic behaviors of (anti-)quark PDFs in the far $x\to 1$ limit. On the side of the upper $\chi^2_R$, we show only the candidate parametrizations that have only a modestly higher $\chi^2_R$ compared to $\chi^2_0$ of the nominal fit. Namely, we show the fits whose $\chi^2_R$ is less than $\chi^2_0 + \sqrt{2 N_\textrm{pt}}$, i.e., fall within the 68\% CL of the statistical uncertainty on $\chi^2_R$ of the nominal fit. Such candidate solutions are a priori eligible for inclusion into the final $\alpha_s$ average over PDF models, even though they would be included with a lower weight into the model average because of their elevated $\chi^2_R$'s. 

Here we observe a striking behavior: while the $\chi^2$ variations among the candidate fits are sizable in the vertical ($\chi^2$) direction, they have modest horizontal variations, compatible with the size of the $\Delta \chi^2=1$ interval, as far as the best-fit $\alpha_s$ are concerned. The picture immediately reminds of the textbook case of the $\chi^2$ variability due to approximately Gaussian random fluctuations in data \cite{BevingtonRobinson}: when one plots the log-likelihood ($\chi^2$) vs. the parameter of interest, e.g., $\alpha_s$, vertical fluctuations of the resulting parabola fall into the interval $\chi^2_R \leq \chi^2_0 + \sqrt{2 N_\textrm{pt}}$ in $\approx 68\%$ of cases, while about 68\% of the horizontal fluctuations fall within the interval $\Delta \chi^2_R \leq 1$.

In our example, the best-fit $\alpha_s$ from fits with alternative parametrizations fall within the $\Delta \chi^2_R=1$ interval from the scan, reflecting statistical fluctuations in the data at 68\% CL for the fixed data set and PDF functional forms. This interval corresponds to the uncertainty $\delta\alpha_s\approx \pm 0.0005$, and this conclusion holds for other candidate selections of baseline experiments. Namely, while the preferred $\alpha_s$ value moves from $\approx 0.1177$ for the CT25prel to 0.1185 for the CT25 selection of data sets, see the bottom of Table~\ref{Tab:LHCnewData}, the pattern of the parametrization dependence is about the same in these fits as in Fig.~\ref{fig:alphas_scan_290forms}, with the parametrization uncertainty staying at the level of 0.0005.

The conclusion based on the ensemble of nearly 300 PDF parametrizations is then that they give very similar $\alpha_s$ values, showing that the determination of $\alpha_s$ is essentially independent of the chosen PDF parametrization.
If only this source of uncertainty were considered, the uncertainty span of approximately $\pm 0.0005$ about the central value would be smaller than both the PDG uncertainty~\cite{ParticleDataGroup:2024cfk} of $\pm 0.0009$ and errors of order $\pm 0.002$ quoted by dedicated PDF-only fits~\cite{Jimenez-Delgado:2014twa,Alekhin:2017kpj,Alekhin:2018pai,Cridge:2021qfd,Ball:2018iqk,Hou:2019efy,H1:2021xxi}.
Given these observations, the final $\alpha_s$ uncertainty estimate does not include a component arising from PDF parametrizations.

\subsection{...but the systematic uncertainty remains}
\label{sec:results-P2}

\begin{figure}[b]
    \centering
    \includegraphics[width=0.59\linewidth]{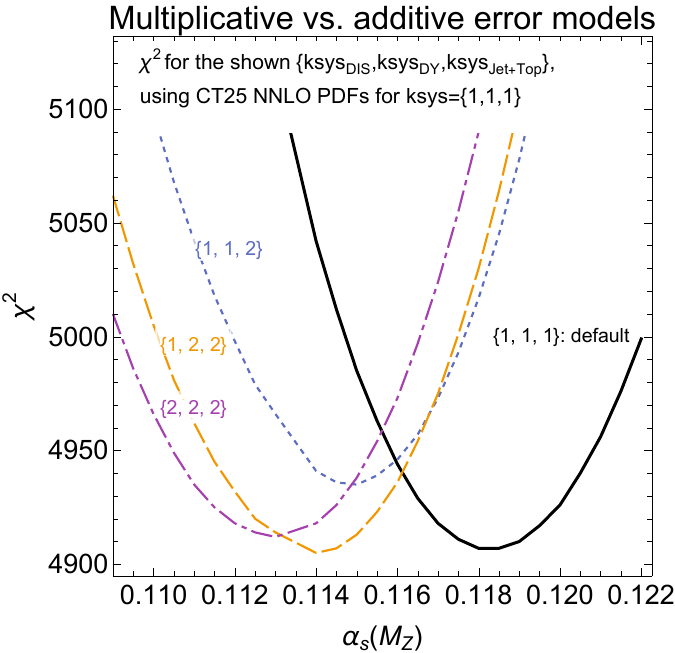}
    \caption{$\chi^2$ profiles for different \texttt{ksys} settings: $\{\texttt{ksys}_{DIS},\texttt{ksys}_{DY},\texttt{ksys}_{\textrm{Jet} + t{\bar t}} \}$ = $\{1,1,1\}$, $\{1,1,2\}$, $\{1,2,2\}$, and $\{2,2,2\}$. These are obtained with a fixed CT25 PDF set determined with the default \texttt{ksys}=1 setting. }
    \label{fig:chi2_for_ksysXYZ}
\end{figure}

The global fit contains two additional notable sources of $\alpha_s$ uncertainties: (1) the dependence on the models of correlated systematic errors discussed in Sec.~\ref{sec:CorrelationLimitations} and (2)
the tensions between preferred values of $\alpha_s$ from the separate categories of data sets.  This section addresses the first aspect and illustrates the possible impact of the choice of the reference normalization factors $X_k$ in the experimental correlation matrices $\beta_{k\alpha}=X_k \sigma_{k\alpha}$, which is controlled in our fit by the parameter \texttt{ksys}. Section~\ref{sec:Tolerances} will follow with addressing the second aspect.
 
Since the experiments normally do not publish their recommended $X_k$, a variety of ``guesses'' about $X_k$ can be, and have been, made in the past PDF fits \cite{Gao:2013wwa,Ball:2012wy}. As illustrative possibilities, we consider two such prescriptions, which we label as ``\texttt{ksys=1}'' and ``\texttt{ksys=2}''.
According to the \texttt{ksys=1} prescription, each correlated systematic element $\beta_{k\alpha}$ (initially provided as a percentage fraction $\sigma_{k\alpha}$) is normalized to the current theory prediction $T_k$ for each data point. This treatment is more suitable when the corresponding experimental error is multiplicative. Major systematic errors for the LHC experiments, for example jet energy scale systematic errors, are treated as multiplicative by the experiments.\footnote{The luminosity error of LHC inclusive jet cross sections is treated as additive.} On the other hand, this prescription can introduce a theory bias in the effect of systematic errors, reflected e.g. in the dependence on the version of the ``$t$'' method for normalization of correlated errors seen in the past analyses \cite{Gao:2013xoa}. The majority of fits by CTEQ-TEA and other groups choose \texttt{ksys=1} for all correlated errors as a default. 

The \texttt{ksys=2} prescription, instead, normalizes the correlated systematic element to the experimental central value
$D_k$. It is most suitable for normalizing additive errors. It can potentially lead to a large D'Agostini bias~\cite{DAgostini:1993arp,DAgostini:1999gfj,Ball:2009qv} (characteristically producing a downward bias in the best-fit cross section) when the underlying error is multiplicative.

For the ensuing discussion, we divide our baseline data set of measurements into three subsets: 1) deep-inelastic scattering (DIS), 2) Drell-Yan (DY) pair production, 3) inclusive jet and $t\bar t$ production. In each subset, we normalize all $\beta_{k\alpha}$ as though they are multiplicative (\texttt{ksys=1}) or additive (\texttt{ksys=2}), with the caveat that the \texttt{ksys=1} option is more likely to be closer to the truth. Measurements do contain a mix of multiplicative and additive errors. The true value of $\alpha_s$ extracted from the global fit is somewhere between the extremes corresponding to the ``all \texttt{ksys=1}'' and ``all \texttt{ksys=2}'' options, but likely closer to \texttt{ksys=1}. This setup allows us to explore sensitivity to systematic error modeling, particularly for hadron jet production observables that exhibit a very large sensitivity.

\paragraph{Impact of systematic-error normalizations for fixed PDFs.} We first demonstrate what happens in the final CT25 fit when we change the default setting of using ``ksys=1'' in all three measurement categories, labeled as  $\{\texttt{ksys}_{DIS},\texttt{ksys}_{DY},\texttt{ksys}_{\textrm{Jet} + t{\bar t}} \}$ = $\{1,1,1\}$.  First, to illustrate how the \texttt{ksys} choice modifies $\chi^2$ when all other settings are unchanged, let us examine the $\chi^2$ dependence for several combinations of \texttt{ksys} values \textit{while using the same PDF ensemble}. Figure~\ref{fig:chi2_for_ksysXYZ}
illustrates these changes, by showing $\chi^2$ vs $\alpha_s$ computed with 
$\{1,1,1\}$, $\{1,1,2\}$, $\{1,2,2\}$, and $\{2,2,2\}$ \texttt{ksys} choices, each using the same nominal CT25 data set and best-fit PDFs obtained with $\{1,1,1\}$.

Both the minimal $\chi^2$ and best-fit $\alpha_s$ values change according to the \texttt{ksys} choices, covering the ranges of 4880-4530 and 0.1155-0.1185, respectively. The minima of the parabolas in this case correspond to the following $\alpha_s$ values: $\{1,1,1\} \rightarrow 0.1185$, $\{1,1,2\} \rightarrow 0.1150$, $\{1,2,2\} \rightarrow 0.1140$, and $\{2,2,2\} \rightarrow 0.1130$. The change from \texttt{ksys}=1 to 2 for jet+$t\bar t$ data sets alone incurs a significant downward $\alpha_s$ shift, which is further amplified by using \texttt{ksys}=2 for the other two groups of processes.

\begin{figure}[tbp]
\centering
\includegraphics[width=0.79\linewidth]{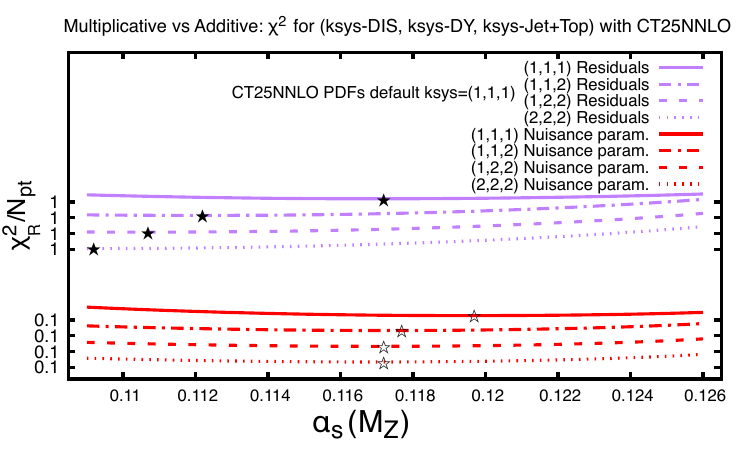}
\caption{Separate contributions to $\chi^2_{R}/N_\textrm{pt}$ from the best-fit summed data residuals and nuisance parameters, as exemplified in Eq.~(\ref{minchi2}), for the four \texttt{ksys} combinations shown in Fig.~\ref{fig:chi2_for_ksysXYZ}.  The theory predictions are obtained with the CT25NNLO PDFs determined using \texttt{ksys}=\{1,1,1\}.}
\label{fig:tot-chi2-ppa21b02}
\end{figure}

It is interesting to investigate further into the details of the fits to understand the driving mechanism of these shifts. First, it can be noted that, as far as the $\alpha_s$ dependence goes, the data-point residuals are more responsive to the form of the systematic errors than the best-fit nuisance parameters. This can be seen from Fig.~\ref{fig:tot-chi2-ppa21b02} that separately plots the contributions from the best-fit residuals and nuisance parameters, given respectively by two sums on the right-hand side of Eq.~(\ref{minchi2}), from the same four $\alpha_s$ scans with the fixed  $\{1,1,1\}$ PDF as in Fig.~\ref{fig:chi2_for_ksysXYZ}. The curves from the four scans are stacked on top of one another so as to fit in the figure. Stars indicate the minima of their corresponding parabolas.  Positions of the stars for residuals can go as low as $\alpha_s(M_Z)=0.109$ for $\{2,2,2\}$. The stars for the nuisance parameters stay between 0.117 and 0.120.

\begin{figure}[htbp]
    \centering
    \includegraphics[width=0.49\linewidth]{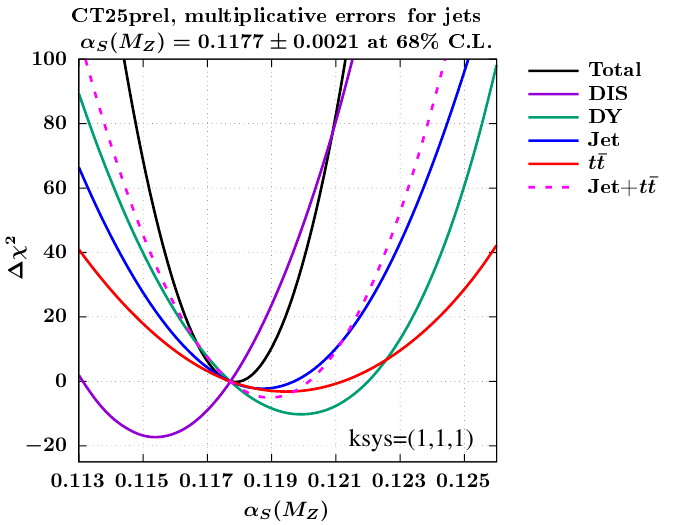}
    \includegraphics[width=0.49\linewidth]{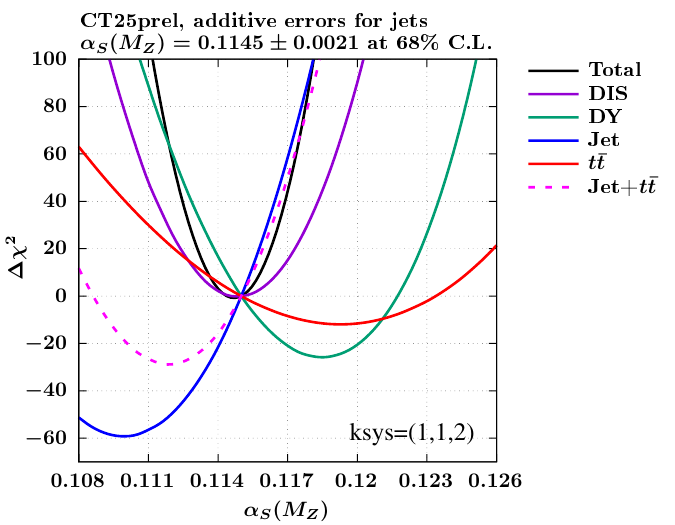}
    \caption{Left: $\Delta\chi^2$ profiles vs. $\alpha_s(M_Z)$ for
    \texttt{ksys=1} applied to all data sets. Right: same, with \texttt{ksys=2} applied to jet data sets, \texttt{ksys=1} to the rest. Curves are shown for the full-baseline global fit (black), DIS (purple), DY (green), jets (blue), $t\bar{t}$ (red), and jet+$t\bar{t}$ (dashed magenta). Each parabola is obtained from the same global fit result, projected onto the corresponding data subset.}
    \label{fig:alpha_s_chi2_scan}
\end{figure}

\begin{figure}[htbp]
    \centering
    \includegraphics[width=0.49\linewidth]{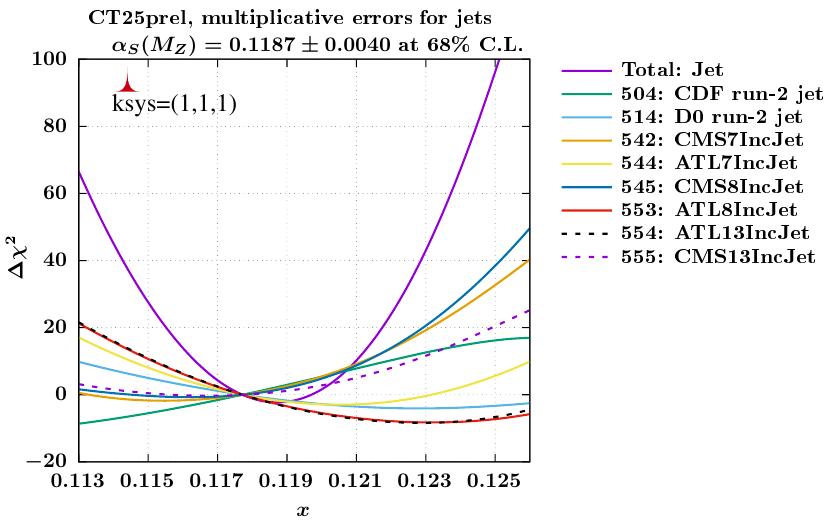}
    \includegraphics[width=0.49\linewidth]{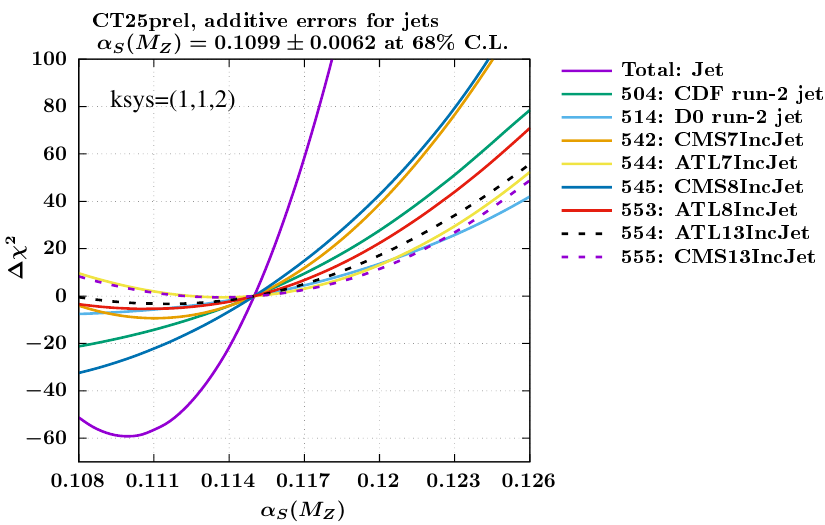}
    \caption{$\Delta\chi^2$ profiles for individual jet-production measurements only in the $\alpha_s$ scans with the \texttt{ksys=1} (left) and \texttt{ksys=2} (right) prescriptions for jet data sets. The $\alpha_s(M_Z)$ values in the plot headers are for the jet data subset only.}
    \label{fig:alphaSJetKsys}
\end{figure}

\paragraph{Impact of systematic-error normalizations with fitted PDFs.}
Figures~\ref{fig:alpha_s_chi2_scan} and \ref{fig:alphaSJetKsys} depict this dynamics in the complete $\alpha_s$ scans, now done using the CT25prel baseline data set that prefers a slightly lower $\alpha_s(M_Z)<0.118$.  In these figures, the PDFs are fitted for the indicated \texttt{ksys} and $\alpha_s$.

Figure~\ref{fig:alpha_s_chi2_scan} illustrates how the $\chi^2$'s for groups of data sets in the CT25prel baseline vary in an $\alpha_s$ scan with $\texttt{ksys}=\{1,1,1\}$ in the left subfigure and $\{1,1,2\}$ in the right one. The curves depict $\Delta\chi^2 \equiv \chi^2_R(\alpha_s)-\chi^2_{R,0}$ vs. $\alpha_s$ -- the changes in $\chi^2_R$ compared to a reference value $\chi^2_{R,0}$ -- for six different groups of data sets in the CT25prel fit: the full baseline (black line), DIS (purple), Drell--Yan pair (green), inclusive jet (blue), and $t\bar{t}$ production (red), as well as the combined jet+$t\bar{t}$ data (dashed magenta). All curves are obtained in one scan in each subfigure. The reference value, $\chi^2_{R,0}$, is the minimum $\chi^2_R$ with respect to both $\alpha_s$ and PDFs. By construction, the $\Delta\chi^2$ curves for all groups of processes intersect (are equal to zero) at the best-fit $\alpha_s$ of the global fit. The minima of the parabolas represent the preferred $\alpha_s$ values for each group of processes. 

With the CT25prel baseline, the default global fit, using the \texttt{ksys=1} treatment (left panel), gives  
\begin{equation}
\alpha_s(M_Z) = 0.1177 \pm 0.0021 \quad \text{(68\% CL)},
\end{equation}
where the 68\% CL is computed according to $\Delta \chi^2=37$, the increase in the global $\chi^2$ consistent with the CT18 tolerance definition. This $\alpha_s$ value is in good agreement with the current world average. 
At the same time, we observe that the preferred $\alpha_s$ values vary across the processes. For example, the DIS experiments prefer the lowest value ($\alpha_s(M_Z) = 0.1152$).\footnote{In the CT25 fit, the majority of DIS experiments prefer $\alpha_s(M_Z)<0.118$, but this preference is especially pronounced for the HERA DIS data at $Q<3.5$ GeV and BCDMS $F_2^p$ data.} Drell--Yan pair and top-quark pair production prefer higher values, 0.1198 and 0.1194, respectively. Inclusive jet production prefers an intermediate value ($\alpha_s(M_Z) = 0.1187$). The spread of preferences in Fig.~\ref{fig:alpha_s_chi2_scan} (left) reflects the differing pulls on $\alpha_s$ from various experiments and highlights tensions in the data.

When the \texttt{ksys=2} prescription is applied to the jet data only in Fig.~\ref{fig:alpha_s_chi2_scan} (right), the global minimum of total $\chi^2_R$ prefers a much lower value, 
\begin{equation}
\alpha_s(M_Z) = 0.1145 \pm 0.0021 \quad \text{(68\% CL)}.
\end{equation}
In this scenario, the ordering of preferred values is markedly different: jet production measurements favor by far the smallest value ($\alpha_s(M_Z) =0.1099$), $t\bar t$ production still pulls toward the largest value ($\alpha_s(M_Z)=0.1191$), and DIS ($\alpha_s(M_Z)=0.1146$) and Drell--Yan process ($\alpha_s(M_Z)=0.1187$) lie in-between. 
Compared to the \texttt{ksys=1} case, 
the pull from jet-production measurements to a significantly lower value of $\alpha_s$ is accompanied by an equally significant reduction in the jets' $\chi^2$, which reflects strong sensitivity of these measurements to the treatment of correlated systematics.

To further investigate the differences between the \texttt{ksys=1} and \texttt{ksys=2} scenarios in the case of inclusive-jet production measurements (in the CT25prel fit), Fig.~\ref{fig:alphaSJetKsys} separately illustrates the $\Delta \chi^2$ profiles of individual jet experiments. As before, the curves have $\Delta \chi^2=0$ at the best-fit $\alpha_s$ of the respective global fit. The \texttt{ksys=1} case is shown in the Fig.~\ref{fig:alphaSJetKsys} (left), where the D\O{} Run-2 and ATLAS  measurements at 7, 8, and 13~TeV collision energies prefer relatively large $\alpha_s$, while the CDF Run-2 and CMS measurements at 7, 8, and 13~TeV prefer smaller values. Their interplay produces a combined $\alpha_s$-minimum from jet production at
\begin{equation}
\alpha_s(M_Z) = 0.1187 \pm 0.0040 \quad \text{(68\% C.L.)},
\end{equation}
consistent with the world average.

By contrast, when the \texttt{ksys=2} treatment is applied to jet data sets in  Fig.~\ref{fig:alphaSJetKsys} (right), all of them prefer a much smaller $\alpha_s$. The CMS 8~TeV and CDF Run-2 data especially pull toward smaller $\alpha_s$ values, compared to the preference of D\O{} and ATLAS for a somewhat larger $\alpha_s$. This produces a much lower central value
\begin{equation}
\alpha_s(M_Z) = 0.1099 \pm 0.0062 \quad \text{(68\% C.L.)},
\end{equation}
that is well below the world average and with a larger uncertainty. 

In summary, the comparison between the two different treatments for the correlated systematic uncertainties in inclusive jet production, \texttt{ksys=1} and \texttt{ksys=2}, demonstrates that these measurements are particularly sensitive to the normalization of systematic errors.\footnote{An analogous sensitivity has been noted in the NNPDF precision determination \cite{Ball:2025xgq} of $\alpha_s(M_Z)$, whereby a fit with a floating $\alpha_s$ in the covariance matrix produces a higher value than the final result consistent with the additive prescription (the analog of our \texttt{ksys}=2).} As we emphasized, requiring all correlated errors to be normalized according to the \texttt{ksys=2} (additive) choice may exaggerate the downward pull on $\alpha_s$; however, other (theoretical) biases may be present also with the \texttt{ksys=1}  prescription. 
Given that the major experimental errors in jet production are regarded as multiplicative, this exercise may exaggerate the possible effect, but still serves as a useful reminder of a vulnerability in the standard publication format for systematic errors.


\section{Uncertainty estimation: tolerances \label{sec:Tolerances}}

\subsection{The \texorpdfstring{$\Delta \chi^2=1$}{Delta chi-squared=1} criterion \label{sec:DeltaChi2Eq1}}

The behavior of $\chi^2$ depicted in Sec.~\ref{sec:results-P2} does not yield a single answer for the uncertainty on $\alpha_s$ compatible with the CT25 global data. One of the conclusions from Sec.~\ref{sec:results-P2} is that, regardless of the implementation of the experimental correlation models, we observe a larger-than-normal spread among $\alpha_s$ preferred by different classes of experiments (DIS, DY, and Jet and top-quark pair production (Jet+$t{\bar t}$) combined), as quantified by the positions of the minima of their $\chi^2$ profiles in Fig.~\ref{fig:alpha_s_chi2_scan}. Consequently, several uncertainty estimates are reasonable. 

We first observe that the dependence of the total $\chi^2$ observed in our $\alpha_s$ scans, such as the one showed by the solid line in Fig.~\ref{fig:alphas_scan_290forms}, is highly consistent with a parabolic fit, with the logarithmic derivative $d\ln{\Delta\chi^2_R}/d(\ln{\alpha_s-\alpha_{s,0}})$ staying in the range 1.9-2.2 over the interval $\alpha_s(M_Z)=0.110-0.125$ covered in the CT25 scan. Based on the total $\chi^2$, the standard deviation is estimated as $\delta_\textrm{68\%} \alpha_s(M_Z) \sim 0.0004$ either from the quadratic fit 
$\chi^2=6.695\cdot 10^6\ \left(\alpha_s(M_Z) -0.1183\right)^2$ or from the condition $\Delta\chi^2 =1$ applicable for a single-parameter probability distribution \cite{BevingtonRobinson}.

The $\Delta \chi^2=1$ uncertainty is interpreted as the frequentist 68\% central confidence interval on the ``$\alpha_s$ parameter'' determined from imaginary repetitions (replicas) of the same experiments using the same fitting methodology. Each such replica consists of the same selection of data points, which differ only in their random fluctuations within Gaussian uncertainties of the input measurements.

But, this surrogate parameter may differ from the true $\alpha_s$ realized in nature if the input data are inconsistent. Such tensions reveal non-negligible data selection uncertainty, which is a part of the model (epistemic) uncertainty. 

Not only is the $\Delta \chi^2=1$ uncertainty $\approx 0.0004$ smaller the world average quoted in the PDG~\cite{ParticleDataGroup:2024cfk}; by such measure the discrepancies among the input $\alpha_s$ values from different classes of processes are truly significant. In particular, we noticed that the DIS measurements show a large disagreement with DY and Jet+$t\bar{t}$ measurements. This behavior reflects the presence of tensions within the CT25 global data set, which can also be seen at the level of individual data sets. 

When theory perfectly agrees with data, the $\chi^2_E/N_\textrm{pt}$ ratios of individual experiments fluctuate around the expectation of one with a known probability. What is actually seen in all global fits is that the spread of $\chi^2_E/N_\textrm{pt}$ is wider than this normal distribution arising from purely stochastic fluctuations. 
One can demonstrate this enlargement of discrepancies by considering histograms of effective Gaussian variables $S_E$ constructed from $\chi^2_E$ \cite{Lai:2010vv,Hou:2019efy}.
While, in the absence of tensions, the distribution of $S_E$ is consistent with the standard normal distribution $\mathcal{N}(0,1)$, the empirically obtained histograms of $S_E$ for the CT, MSHT, and NNPDF global fits are wider than $\mathcal{N}(0,1)$ \cite{Kovarik:2019xvh}. 

In this and the following section, we compare several strategies to account for the data selection uncertainty. First, we focus on traditionally used methods that increase the input uncertainties or introduce tolerance on the final uncertainty to address incomplete knowledge of the underlying model. Then, we apply Bayesian methods to tackle mutual consistency of the individual $\alpha_s$ determinations. 
\begin{enumerate}
    \item Section~\ref{sec:PDG-Uncertainty-combination} applies the Particle Data Group (PDG) prescription for combining inconsistent measurements.  
    \item Sections~\ref{sec:Global-Dynam} and \ref{sec:DTproperties} review the global and dynamical tolerance prescriptions; in Sec.~\ref{sec:ClusteringSafety}, we explore stability of the dynamic tolerance with respect to clustering of data points.
    \item Section~\ref{sec:reclustering} explores the impact of data reclustering on the $\alpha_s$ uncertainty.
    \item Section~\ref{sec:BayesianModels} estimates the $\alpha_s$ uncertainty using a Bayesian hierarchical model and Bayesian model averaging.
\end{enumerate}

\subsection{The Particle Data Group prescription}
\label{sec:PDG-Uncertainty-combination}
 
In order to quantify the level of tension, we first consider a simple combination of
the $\alpha_s$ central values and uncertainties obtained from the three classes of experiments analyzed in Sec.~\ref{sec:results-P2}, namely DIS, DY, and Jet+$t\bar t$ combined. We analyse $\Delta \chi^2_R$ profiles for the CT25 baseline, like those in Fig.~\ref{fig:alpha_s_chi2_scan}, and treat the $\alpha_s$ values corresponding to the minima of the individual profiles as independent determinations.\footnote{Although the $\chi^2$ profiles here are extracted from a global fit, we argue that this approximation is reasonable for the present analysis.} 
We find that the associated uncertainties for each class of experiments are nearly Gaussian as well, since each $\chi^2$ profile  for the $i$-th class is well approximated by a quadratic function
\begin{equation}
\chi^2 (\alpha_s,\alpha_{s,i},\sigma_i) =  \frac{(\alpha_s - \alpha_{s,i})^2}{\sigma_i^2}\ ,
\label{eq:chi2_alpha_ind}
\end{equation}
with free parameters $\alpha_{s,i}$ and $\sigma_i$ denoting the central value and uncertainty. 
Table~\ref{tab:gaussian_combination} lists preferred $\alpha_{s,i}$ and $\sigma_i$, as well as their combination given by the weighted average. 
\begin{table}[tb]
\centering
\begin{tabular}{|l|c|c|c|c|}
\hline
 & DIS & DY & Jet+$t\bar{t}$ & Weighted average \\
\hline
$\alpha_{s,i}(M_Z) \times 10^3$ & 115.14 & 118.62 & 120.65 & \textbf{118.37} \\
Error $(\sigma_i \times 10^3)$ & 0.629 & 0.743 & 0.543 & \textbf{0.360} \\
\hline
\end{tabular}
\caption{Individual and combined determinations of $\alpha_s(M_Z)$ from DIS, DY, and Jet+$t\bar{t}$ data for the CT25 NNLO baseline. Errors in the bottom row are obtained with $\Delta\chi^2 = 1$.}
\label{tab:gaussian_combination}
\end{table}
The combination value $\overline\alpha_s = 0.1184\pm 0.0004$ is consistent with the respective direct determination from total $\chi^2$ in Sec.~\ref{sec:DeltaChi2Eq1}. Again, the resulting uncertainty is significantly smaller than for the current world average~\cite{ParticleDataGroup:2024cfk}. More importantly, the level of tension between the input data classes is characterized by the large value of $\sum_i\chi^2(\overline\alpha_s,\alpha_{s,i},\sigma_i) \approx 44$ for (3-1) degrees of freedom (d.o.f.), indicating a dramatically bad quality of the fit.

As such disagreement is unlikely to occur randomly, the Introduction of the Review of Particle Physics \cite{ParticleDataGroup:2024cfk} describes a corrective approach that inflates the input uncertainties or, equivalently, defines the standard deviation to correspond to $\Delta\chi^2=T^2$ with $T^2>1$. 
%
The error-inflation approach introduces a 
scale factor $e_{S_{\text{PDG}}}$  that increases the uncertainties on each individual measurement  so that the reduced $\chi^2/\text{d.o.f.} \approx 1$. This condition is guaranteed by choosing the scale factor to be
\begin{equation}
 e_{S_{\text{PDG}}} = \sqrt{\chi^2/\text{d.o.f.}}\ .
\end{equation}
For the combination of values in Table~\ref{tab:gaussian_combination}, one finds $e_{S_{\text{PDG}}} \approx \sqrt{22} \approx 4.7$. 
In the context of one-dimensional data combination, this condition is equivalent to adopting a global tolerance of $T \simeq 4.7$. The application of the PDG rescaling results in
\begin{equation}
 \alpha_s = 0.1184 \pm 0.0017 \quad (\text{scaled errors, } e_{S_{\text{PDG}}} \approx 4.7).
\label{eq:pdg_uncertainty}
\end{equation}

The PDG prescription rests on the premise that the nominal uncertainties must be increased when tensions suggest possible mismodeling. It is important to note that, as explicitly recommended by the PDG, such a democratic rescaling of all input uncertainties is not advised especially for fits of poor quality, when a more informed procedure should be used. For example, one could try to introduce an additional systematic error contributing only along the parameter direction affected by the tension \cite[Sec.~IV.J in Ref.][]{Kovarik:2019xvh}. 

\subsection{Global and Dynamical Tolerance Criteria}
\label{sec:Global-Dynam}

Nevertheless, the PDG prescription captures the broad idea that, when the control of the underlying model is in doubt, a more conservative estimate linked to testing of the models, i.e., the goodness-of-fit criterion in the frequentist paradigm, may be warranted. The tolerance prescriptions developed by the PDF analyses implement this idea. 

The simplest global tolerance \cite{Pumplin:2001ct} uniformly associates the 68\% CL uncertainty with an increase $\Delta \chi^2=T^2 > 1$ along any direction in space of free parameters in the Hessian formalism. This prescription may be too simplistic, as e.g. tensions among experiments are likely to constrain different combinations of parameters in dissimilar ways. Sometimes one experiment may dominate the constraint along an orthogonal direction (an eigenvector), but more often it is a combination of experiments. 
The dynamic tolerance, which has been developed in several forms, leverages constraints from individual experiments to obtain $T^2$ values independently for each orthogonal direction in the parameter space. In this study, we examine tolerance along the direction corresponding to the $\alpha_s$ free parameter. We also discuss how the dynamic tolerance relates to the global and $\Delta \chi^2=1$ ones under reclustering of input data. 

{\bf Global tolerance (GT).} A common way to introduce either the global or dynamic tolerance is to consider quantiles of the probability density $P(x)$ supported on $-\infty < x < \infty$. The quantile $\xi(P,v)$ at cumulative probability $v$ is defined by 
\begin{equation}
    \int_{-\infty}^{\xi(v)} P(x) dx = v.
\label{quantile}
\end{equation}
One might consider $P(x)$ to be the $\chi^2$ distribution of the baseline data set. Then, one determines the global tolerance $T^2$, or equivalently constructs the 68\% CL interval $\alpha_{s,\min} \leq \alpha_s(M_Z) \leq \alpha_{s,\max}$, from the quantile $\xi(\chi^2(N_\textrm{pt}),0.68)$ of the $\chi^2$  distribution with $N_\textrm{pt}$ degrees of freedom. This is a more general procedure to compute $T^2$ than the PDG-inspired estimate in Sec.~\ref{sec:PDG-Uncertainty-combination}, as it operates directly with $P(x)$ instead of effective data points. In practice, other values of $T$ may be implemented, such as the GT equal to the average of $T$ values along independent parameter directions in the DT prescription.

{\bf Dynamical Tolerance (DT).} When the quantiles in Eq.(\ref{quantile}) are computed for each $\chi^2_{E}$ profile of the individual data sets, and then a single uncertainty range is constructed from their combination, we refer to it as the ``dynamical tolerance uncertainty''~\cite{Martin:2009iq,Kovarik:2019xvh}. Suppose the GT applied just to experiment $E$ renders the uncertainty range $\alpha_{s,\min}^{(E)} \leq \alpha_s(M_Z) \leq \alpha_{s,\max}^{(E)}$. The DT interval for the full data set is constructed separately from the upper and lower boundaries for each experiment as 
\begin{equation}
\alpha_{s,\min}^\textrm{DT}  \equiv \max_{E}\alpha_{s,\min}^{(E)},\quad \alpha_{s,\max}^\textrm{DT}  \equiv \min_{E}
\alpha_{s,\max}^{(E)}.   
\label{max-min-err}
\end{equation}

{\bf The two-tier CT18 uncertainty.}  The CT \cite{Lai:2010vv,Gao:2013xoa} and MSHT \cite{Martin:2009iq} groups implement variations of this generic procedure. The global and dynamical tolerance can be combined into a more complex constraint, such as the two-tier tolerance prescription used by CT10, CT14, and CT18 PDFs. In this prescription, the global tolerance with $T^2 \sim 37$ is combined with the ``tier-2'' dynamical tolerance at the 68\% CL. The goal is of the current study is to bypass this definition, which otherwise would require, in particular, to estimate the global $T^2$ through the quantile argument or by another means. 

{\bf A simplified tolerance.} We will approximate both the CT and MSHT procedures by the following simplified prescription based on the effective Gaussian variable $S$. 

$S$ represents a type of a $Z$ score, i.e., it is a function of $\chi^2$ that is distributed approximately according to $\mathcal{N}(0,1)$ and is easier to work with than the $\chi^2$ distribution with $N_\textrm{pt}$ degrees of freedom. For each $\chi^2$, the corresponding $S(\chi^2, N_\textrm{pt})$ can be computed using either a simpler Fisher's approximation \cite{Fisher:1925},
\begin{equation}
S(\chi^2,N_\textrm{pt}) = \sqrt{2\chi^2(N_\textrm{pt})} - \sqrt{2 N_\textrm{pt}-1} \,,     
\label{Fisher}
\end{equation}
or a more precise formula by T. Lewis \cite{Lewis1988},
\begin{equation}
    S(\chi^2,N_\textrm{pt}) = \left(- 1 + \frac{1}{9\,N_{\textrm{pt}}}+ 
\frac{1}{1 - \tfrac{1}{6}\,\ln\!\left(\frac{\chi^{2}}{N_{\textrm{pt}}}\right)}
\right)
\frac{\sqrt{18\,N_{\textrm{pt}}}}{1 + \tfrac{1}{18\,N_{\textrm{pt}}}}.
\label{Lewis}
\end{equation}

For both the CT and MSHT tolerance implementations, which are fully
specified in their respective articles, the following mapping holds for the credibility interval 
up to subleading terms in $1/N_\textrm{pt}$:
\begin{equation}
    \alpha_{s,\min} \leq \alpha_s \leq \alpha_{s,\max} \quad \Leftrightarrow \quad S(\chi^2(\alpha_s), N_\textrm{pt}) \leq S(\chi^2_0, N_\textrm{pt}) + \xi(\mathcal{N}(0,1), v).
    \label{SDTol}
\end{equation}
That is, the uncertainty interval at probability $v$ is found by selecting those $\alpha_s$ values for which the Gaussian variable $S(\chi^2(\alpha_s), N_\textrm{pt})$ does not exceed its value $S(\chi^2_0, N_\textrm{pt})$ at the global minimum by more the quantile $\xi(\mathcal{N}(0,1), v)$ determined from $\mathcal{N}(0,1)$. 

The prescription (\ref{SDTol}) simplifies the calculation of the probability intervals, while being numerically close to the CT and MSHT full implementations. To compute uncertainties at 68\% and 90\% CL, one takes $\xi(\mathcal{N}(0,1), 0.68)=0.468$ and $\xi(\mathcal{N}(0,1), 0.90)=1.28$.

\begin{figure}[tb]
\centering
\includegraphics[width=0.8\linewidth]{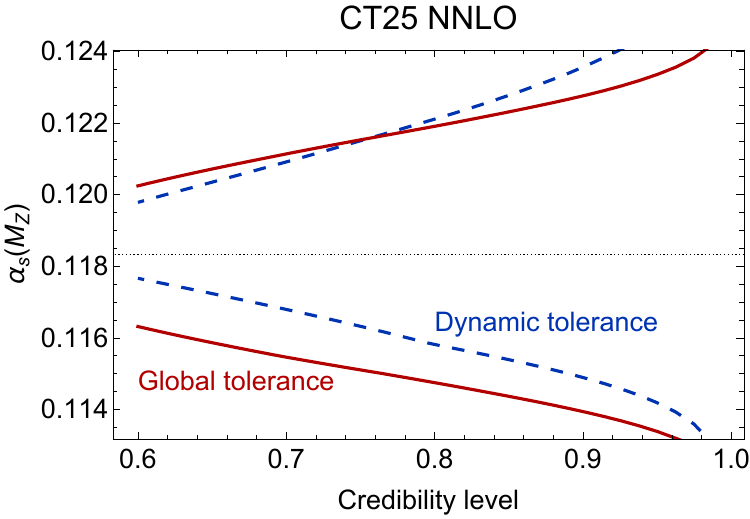}
\caption{$\alpha_s{(M_Z)}$ uncertainties in the CT25 NNLO fit as a
  function of the credibility level for the global and dynamical
  tolerance criteria. The DT error band is computed using 49 published
data sets of the CT25 baseline.}
\label{fig:alps-vs-CredLev}
\end{figure}

\begin{figure}[tb]
\centering
\includegraphics[width=0.48\linewidth]{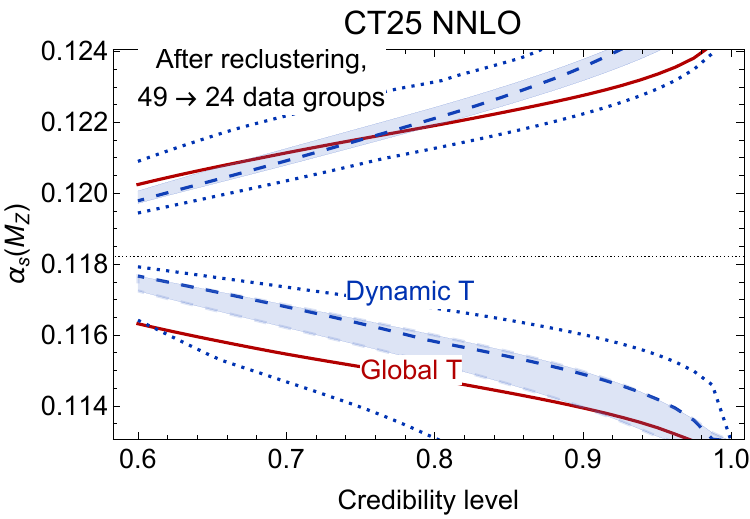}\quad
\includegraphics[width=0.48\linewidth]{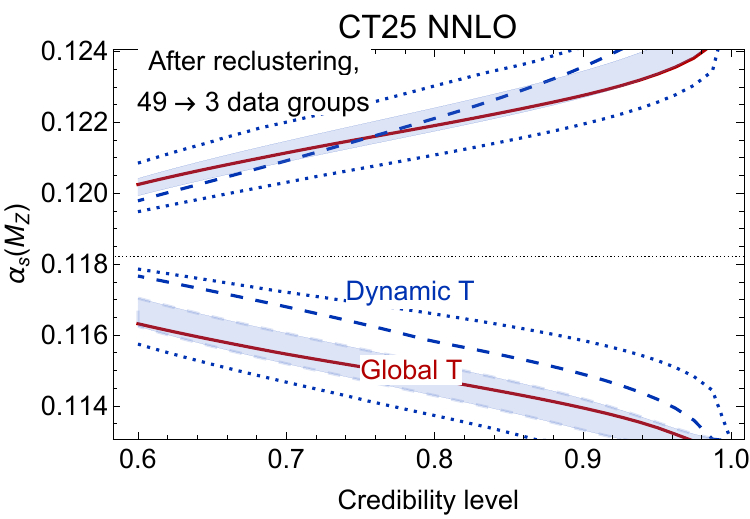}
\caption{Dependence on the dynamic tolerance uncertainties on the
  clustering of experiments in the CT25 NNLO fit. The red solid and
  blue dashed curves indicate the same GT and DT uncertainties as in 
  Fig.~\ref{fig:alps-vs-CredLev}. On the top we superimpose DT
  uncertainties from 10000 alternative clusterings of experiments in
  which the 49 original data sets were randomly combined in 24
  clusters (left) and 3 clusters (right). At each credibility level,
  6800 instances of the revised DT uncertainties fall within the
  light-blue bands. Blue dotted lines indicate the most extreme DT
  uncertainties
  among the 10000 instances. On average, the span of the DT errors reduces with more clusters and approaches the GT error with fewer clusters. }
\label{figs:alps-clustering}
\end{figure}

\subsection{Properties of the dynamic tolerance}
\label{sec:DTproperties}

The dynamical tolerance (DT) has proven to be convenient when accounting for
constraints on PDF uncertainties from individual experiments. That
said, at the detailed level, the DT has a more complicated behavior than the
GT. We illustrate these aspects using the results of the $\alpha_s$ scans with the CT25 and CT25prel baselines described in Sec.~\ref{sec:SourcesOfUncertainties}.

{\bf DT uncertainties neglect correlations among the data sets via the shared PDF parameters.} 
The correlations are considered when obtaining the central prediction and neglected when estimating the
uncertainty. Indeed, the global minimum that defines the central prediction arises from the correlated constraints imposed by all experiments on the PDFs. When computing the quantiles of experiments, 
these correlations are neglected -- this approximation breaks at some level.

{\bf  DT intervals tend to be asymmetric} around the central value given by the position of the global minimum for all experiments.  This can be explicitly seen from Eqs.~(\ref{SDTol}) and (\ref{max-min-err}), showing that the DT intervals are determined from {\it changes} in the probabilities of two most constraining experiments $E$ with respect to the global $\chi^2$ minimum.\footnote{In the complete the CT or MSHT implementations of the DT formulas, such behavior is achieved by applying the quantile to the ratio $S(\chi^2_E)/S(\chi^2_{E,0})$ or $\chi^2_E/\chi^2_{E,0}$, not to $S(\chi^2_E)$ or $\chi^2_E$ itself.}   
For example, Fig.~\ref{fig:alps-vs-CredLev} compares the GT and DT uncertainties in the CT25 NNLO fit for credibility levels
indicated on the horizontal axis. Throughout this article, the GT is found from the quantile of the baseline data set according to Eq.~(\ref{quantile}). The DT is
constructed from the quantiles of 49 contributing data sets and has an obvious asymmetry above and below the central $\alpha_s(M_Z)$. 

 {\bf DT is sensitive to clustering of data.} It depends on what counts as an independent
 experiment. Usually, one constructs the DT it from the $\chi^2$'s of the
 published data sets. It could be equally constructed from some other clusters of data, e.g., 
 by combining (clustering) the published data sets into ``supersets'' or dividing them into ``subsets''.

Figure~\ref{figs:alps-clustering} shows the same default GT and DT uncertainties for the 49 data sets 
as in Fig.~\ref{fig:alps-vs-CredLev}.  In addition, we recomputed the DT 10000 times, every time randomly
combining the 49 data sets into 24 clusters in the left subfigure and into 3
clusters in the right one. For every CL value, we thus quantify the
uncertainty on the DT uncertainty, communicated in the figure by the
filled blue bands containing 6800 instances of the re-clustered DT
intervals, as well as the dotted blue bands (envelopes) indicating the most extreme instances among the 10000 re-clustered DT intervals. From Fig.~\ref{figs:alps-clustering}, it is clear that the average length and uncertainty of the DT interval depend on reclustering. Fewer clusters result in a DT uncertainty closer to the GT uncertainty, corresponding in this example to a single global data cluster. Conversely, finer clustering reduces the DT uncertainty interval. 

Furthermore, we expect that clustering a pair of inconsistent experiments increases the DT interval. Conversely, splitting an inconsistent experiment into two or more clusters may reduce the DT interval.
The reason is that the DT measures the change in the $\chi^2_E$ compared to its value at the best fit. When an experiment $E$ is in a disagreement, the change in its $\chi^2_E$ is steep and often nonlinear. When inconsistent experiments are combined, their combined change may be much slower than for each one individually, resulting in a larger DT error. On the other hand, an internal inconsistency of a data set may be obviated by dividing this data set into incompatible parts, which may reduce the DT error.

\begin{figure}
\centering
\includegraphics[width=0.8\linewidth]{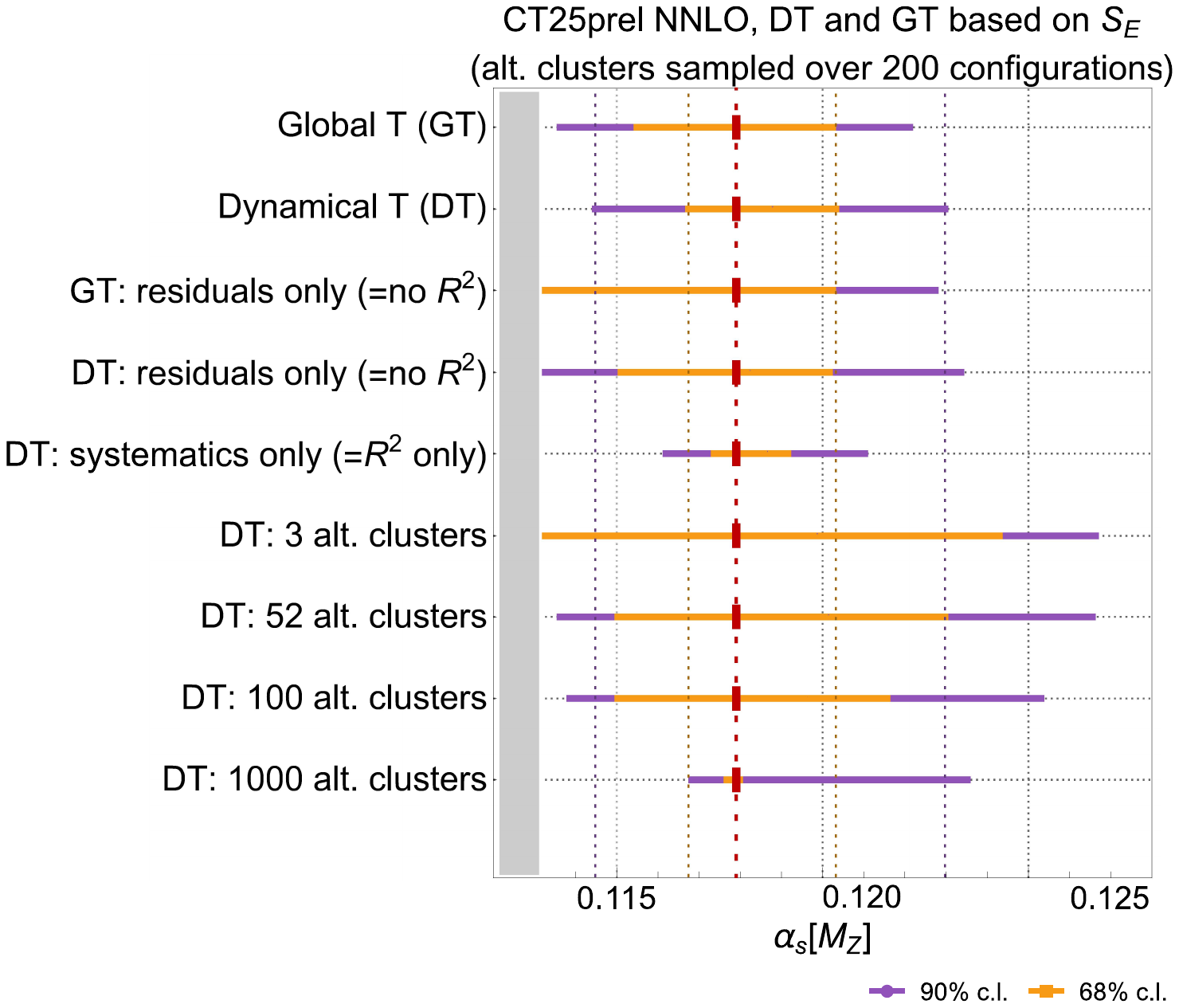}
\caption{Global and dynamical tolerance uncertainties on $\alpha_s(M_Z)$ at the 68\% and 90\% CL for various clustering procedures applied to the CT25prel baseline. DT uncertainties are estimated according to the default prescription, without and with including the sum $R^2$ of best-fit nuisance parameters, and optionally by reclustering the data into 3, 52, 100, and 1000 clusters as explained in Sec.~\ref{sec:reclustering}. For the reclustered data, the DT errors are averaged over 200 alternative configurations. 
}
\label{fig:unc-summary}
\end{figure}

Figure~\ref{fig:unc-summary} summarizes this behavior by showing the
DT uncertainty at 68\% CL and 90\% CL for the CT25prel baseline data
comprised of 52 published data sets. In addition to the GT
uncertainty, corresponding to one cluster containing all data, we show
the DT uncertainty for 52 published data sets, as well as for
alternative clusters (both supersets and subsets)
detailed in Sec.~\ref{sec:reclustering}. 

The figure makes it obvious that the DT uncertainty changes depending
on the clustering procedure. Specifically:
\begin{eqnarray}
  &  \textrm{One cluster\quad}:&\quad  \textrm{DT} \approx \textrm{GT}; \nonumber \\
  &  \textrm{Many clusters}:&\quad \textrm{DT} \to (\Delta \chi^2 =1).
\end{eqnarray}
When all experiments are combined, the dynamical tolerance reduces to the global one.
When data are split into a large number
of clusters (1000 in Fig.~\ref{fig:unc-summary}), the DT uncertainty
essentially reduces to the $\Delta \chi^2=1$ uncertainty, when a single data point (degree of freedom) controls the allowed parameter displacement. 

\subsection{Data clustering safety of uncertainty estimates}
\label{sec:ClusteringSafety}
The previous subsection points out that the DT uncertainty depends on how the data are clustered. The fitting groups typically construct the DT using the full published data sets as independent clusters, but other reasonable clustering choices are obviously possible.

Such sensitivity would raise the concern that the DT uncertainty estimate on a fundamental parameter is {\it unsafe} at some level with respect to the user-chosen data clustering prescription. In contrast, the GT is trivially clustering-safe. 

The key point here is that reclustering of the data sets modifies the
degree of tensions present in the data. For instance, in a version of
the CT25prel fit to 52 data sets shown in Fig.~\ref{fig:unc-summary},
the GT estimate is $\alpha_s(M_Z)=0.1179\pm 0.0024$, i.e. the error is
practically symmetric. The default DT error, formed by the nominal clustering to the 52 data sets, is $0.1179^{+0.0024}_{-0.0011}$, showing a pronounced asymmetry. Upon reorganizing the data sets into 3 clusters, without breaking the data sets, in some cases the data sets are grouped so that their opposite pulls on $\alpha_s(M_Z)$, and in other cases these pulls are maximally exacerbated. When the tensions are largely cancelled, the 3-cluster DT error becomes nearly symmetric and larger than the GT one, e.g., $0.1179^{+0.0032}_{-0.0028}$. When the clustering magnifies the tensions, the 3-cluster DT errors vary more and can be highly asymmetric or not, for instance,  $0.1179^{+0.0021}_{-0.0008}$ or $0.1179^{+0.0016}_{-0.0019}$.

The DT estimates thus in principle should be supplemented by additional tests of the degree of their clustering (un)safety. In light of what was said above, the DT errors for the $\alpha_s$ determination vary between zero and the GT errors. One could take an envelope of these DT errors as a conservative estimate, or an average DT error over many clustering configurations as an optimal estimate. In our specific case, the latter estimate is close to the default one using the 52 individual experiments, and hence the final uncertainty combination will concentrate on the default DT tolerance. 
 
\subsection{Data reclustering and K-folding}
\label{sec:reclustering}
It is interesting to ask whether the tolerance-based methods can quantify the impact of tensions within the available data sets, not only among the data sets. In this subsection, we further investigate this question, which is generally complex but can be elucidated by partitioning the best-fit $\chi^2$ in Eq.~(\ref{minchi2}) into alternative clusters of data residuals $r_{0,k}^{(E)}$ and nuisance parameters (NPs) $\lambda^{(E)}_{0,\alpha}$. This exercise reveals scaling properties of uncertainty estimates with respect to the number of clusters, although the scaling would be further affected by correlations among $r_{0,k}^{(E)}$ that are arguably less impactful and have been neglected. In particular, the estimates for GT and DT based on the residuals only (without $R^2$, the contribution from NPs) are overestimated by assuming $N_\textrm{pt}$ degrees of freedom for the contribution $D^2$ of the residuals instead of the actual $N_\textrm{pt}-N_\textrm{par}$. Working in this framework, we will summarize the procedure employed to partition the published data sets into many clusters illustrated in Fig.~\ref{fig:unc-summary}. Then, we will comment on the potential impact of cross validation procedures in the Monte-Carlo-based approaches on the $\alpha_s$ uncertainty estimate. The reader interested in the Bayesian methods and final combination can proceed directly to Secs.~\ref{sec:BayesianModels} and \ref{sec:combination}.

\subsubsection{Reclustering algorithm \label{sec:reclustering_algorithm}}
If we neglect correlations via the PDF parameters, as was done to justify the DT to start with, then $[r_{0,k}^{(E)}]^2$ and $[\lambda^{(E)}_{0,\alpha}]^2$ from all data sets can be separately and independently collected (``shuffled'') into  alternative clusters of data points, and then the DT can be formed from such clusters. 

Working with the CT25prel baseline, we start by organizing all $\chi_R^2$ components in Eq.~(\ref{minchi2}) into two vectors $\mathbf{v}_r$ and $\mathbf{v}_{\lambda}$ as 
\begin{eqnarray}
\mathbf{v}_r &=& \left\{r^{(1)}_{0, 1}, \dots r^{(1)}_{0, N_\textrm{pt}^{(1)}},
\dots r^{(N_E)}_{0,1},\dots, 
r^{(N_E)}_{0,{N_\textrm{pt}^{(N_E})}}
\right\} \, ,
\label{res-vec}
\\
\mathbf{v}_{\lambda} &=& \left\{\lambda^{(1)}_{0,1},\dots,\lambda^{(1)}_{0,N_\lambda^{(1)}},\dots,\lambda^{(N_E)}_{0,1},\dots,\lambda^{(N_E)}_{0, N_\lambda^{(N_E)}} \right\}  \, ,
\label{R2-vec}
\\
\mbox{so that }\chi^2_R &=& D^2+ R^2, \quad\quad D^2 \equiv ||\mathbf{v}_r||^2 ,  \quad \quad R^2\equiv||\mathbf{v}_{\lambda}||^2\,.
\label{chisq}
\end{eqnarray}
We randomly assign the $\mathbf{v}_r$ and $\mathbf{v}_\lambda$ components to $N_{cl}$ clusters with equal numbers of data points (modulo $N_{cl}$), keeping this assignment same for all $\alpha_s$. We construct the DT according to the same procedure as for the published data sets in Sec.~\ref{sec:Global-Dynam}. Namely, for each cluster $E$ we compute the $\chi^2_E$ quantile at our CL, determine $\alpha_{s,min}^{(E)}$ and $\alpha_{s,max}^{(E)}$ for this quantile, and then construct the DT bounds as in Eq.~(\ref{max-min-err}), where $E$ now runs over the clusters. We examine the $\alpha_s$ dependence of DT by (a) constructing the DT interval for each of $N_{\alpha_s}=21$ values of $\alpha_s(M_Z)$ on a grid between 0.113 and 0.129, and (b) repeating this procedure many (e.g., $>200$)
times, each time reshuffling the assignment of $r_{0,k}^{(E)}$ and/or $\lambda^{(E)}_{0,\alpha}$. We can compute DT for $D^2$ only, $R^2$ only, or the whole $D^2+R^2$. We can do this because, in our approximation, the best-fit residuals and nuisance parameters enter $\chi^2_R$ as
independent degrees of freedom.

In Fig.~\ref{figs:alps-clustering}, the first five lines compare the GT and DT uncertainties at 68\% and 90\% CL computed based on the full $\chi^2_E$, residuals  ($D^2$) only, and nuisance parameters ($R^2$) only. This comparison reveals elevated tension in the $R^2$, causing the DT uncertainty based on $R^2$ only (line 5) to be smaller than those based on the full $D^2+R^2$ (line 2) and $D^2$ only (line 4). Appendix~\ref{app:Reclustering} further explores the GT and DT estimates collected in Fig.~\ref{figs:alps-clustering}, also providing them at other credibility levels.

\subsubsection{K-folding impact on \texorpdfstring{$\alpha_s$}{QCD coupling}}  
\label{sec:K-folding}
Cross validation methods such as jackknife resampling \cite{Quenouille:1949,Tukey:1958,Miller:1974} are widely adopted to estimate the data selection uncertainty. In a classical realization, the parameter of interest is first estimated from a part (fold) of the baseline data; the parameter's expectation value and variance are determined by averaging over multiple random selections of such folds. Cross validation is also a part of MC/ML approaches and involves partitioning the baseline data set into training, testing, and occasionally validating components. One can ask, for instance, to what extent the cross validation technique of ``K-folding'' deployed in the recent NNPDF determination \cite{Ball:2025xgq} of $\alpha_s(M_Z)$ may be sensitive to data tensions. Since K-folding involves reclustering of the baseline data, we will briefly review its possible impacts on the $\alpha_s(M_Z)$ uncertainty.

In the common K-folding procedure, each replica of the data baseline is partitioned into several folds (clusters), 
with one fold set aside as the test data set. The model is then trained on the union of remaining folds. A figure of merit called a ``loss function'' $L$ is optimized so as to achieve good agreement with both the training and testing sets.  The training procedure is iterated by excluding every fold in turn and training on the rest of the folds. In the $\alpha_s$ study, each such iteration  also returns the best-fit $\alpha_s(M_Z)$ and its uncertainty. The final $\alpha_s(M_Z)$ is determined by weighted averaging over all replicas and K-folding iterations. The weights may be given in terms of the respective $\alpha_s$ uncertainties from the iterations.

In the NNPDF study \cite{Ball:2025xgq}, each replica of the published data sets is partitioned into the training set and testing fold as 3:1, without breaking the data sets apart. The $\alpha_s$ uncertainty from each iteration of training is determined essentially according to the $\Delta \chi^2=1$ criterion. The final $\alpha_s(M_Z)$ is obtained by averaging over the data replicas and K-folding iterations using $1/(\Delta\alpha_s(M_Z))^2$ from each iteration as the weight. Their resulting value $\alpha_s(M_Z)=0.1194^{+0.0007}_{-0.0014}$. The lower of these uncertainties is roughly equal to the quadrature sum of our $\Delta \chi^2=1$ and PDF parametrization uncertainties. The NNPDF uncertainty is smaller than our $\delta \alpha_s \approx 0.002$, plausibly because the latter also includes the data selection uncertainty.

We posit, based on the preceding discussion, that the reclustering involved in K-folding should affect $\delta\alpha_s$. For a reliable estimate, $\delta\alpha_s$ should be averaged over a representative ensemble of K-folding/cross-validation partitions.

While such an exhaustive analysis, especially refitting the PDFs for multiple data partitions, is beyond the scope of our article, Appendix~\ref{app:ToyKFolding} highlights several considerations based on the following simplified exercise. 

\begin{figure}[ht]
\centering
\includegraphics[width=0.48\linewidth]{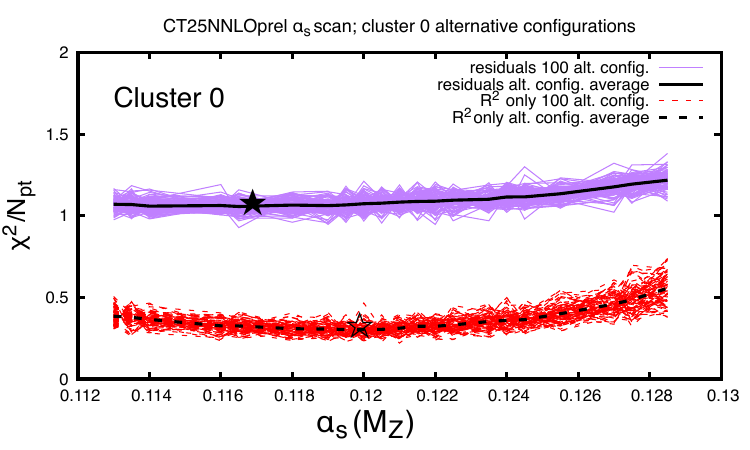}\quad
\includegraphics[width=0.48\linewidth]{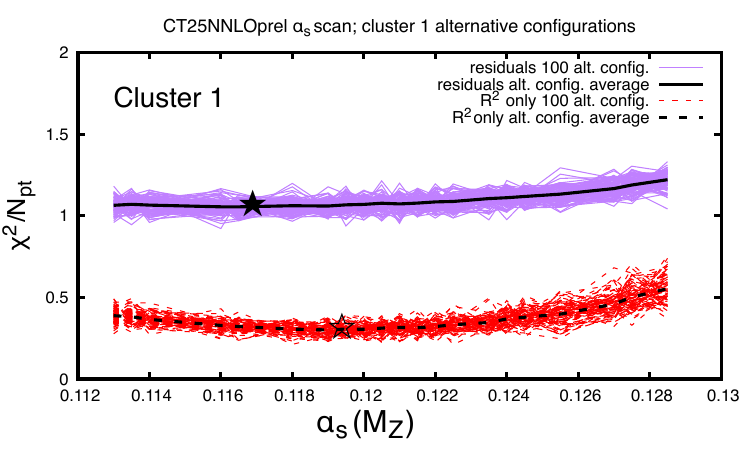}\\
\includegraphics[width=0.48\linewidth]{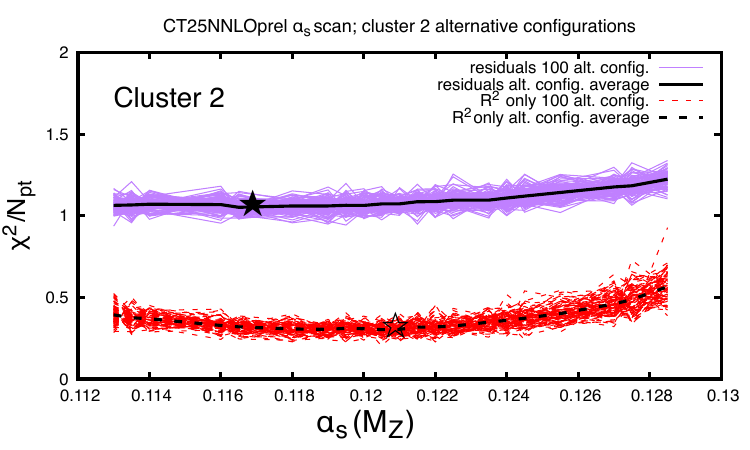}\quad
\includegraphics[width=0.48\linewidth]{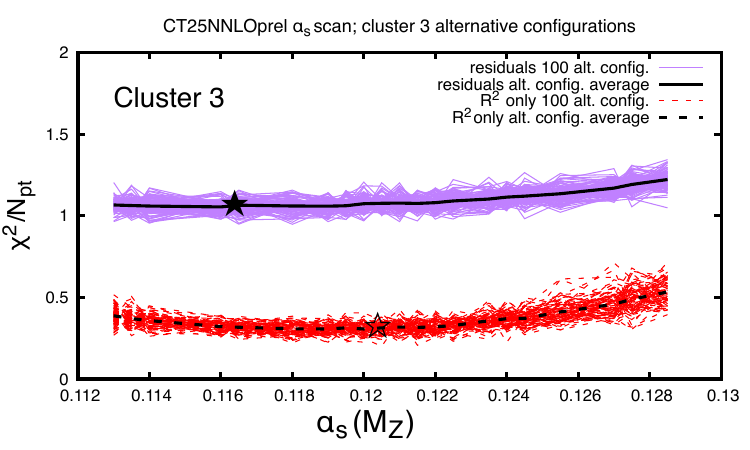}
\caption{$\chi^2/N_\textrm{pt}$ as a function of $\alpha_s(M_Z)$ for 100 alternative configurations of residuals and nuisance parameters partitioned into four equal-sized clusters.}
\label{fig:chi2-vs-alps-K-folds}
\end{figure}

\begin{figure}[ht]
\centering
\includegraphics[width=0.55\linewidth]{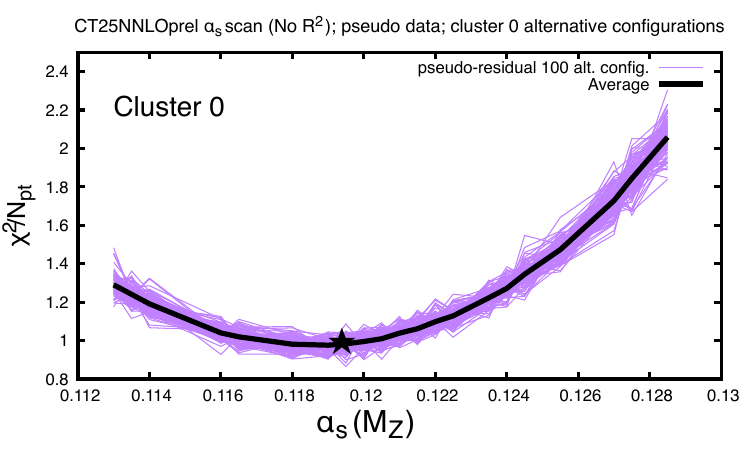}
\caption{$\chi^2/N_\textrm{pt}$ as a function of $\alpha_s(M_Z)$ for cluster 0 with random reshuffling of pseudoresiduals from pseudodata without tensions.}
\label{fig:chi2-vs-alps-pseudo}
\end{figure}

Working again with vector $\mathbf{v}_r$ of about 4000 best-fit residuals from the CT25prel baseline, we partition the residuals into four clusters of about 1000 members each. Similarly, we partition nuisance parameters from vector $\mathbf{v}_\lambda$ into four approximately equal clusters. We repeat this step 100 times, each time randomly modifying the partitions. For each partition (alternative configuration) and for the averages over 100 alternative configurations, we plot the $\alpha_s$ dependence of $\chi^2_R/N_\textrm{pt}$ in each cluster in Fig.~\ref{fig:chi2-vs-alps-K-folds}. We make an important observation, namely, that the $\chi^2_R/N_\textrm{pt}$ distributions in Fig.~\ref{fig:chi2-vs-alps-K-folds} are flatter than would be expected from an ideal distribution of a cluster of $\approx$ 1000 made-up residuals that are consistently distributed according to $\mathcal{N}(0,1)$ around a single reference value of $\alpha_s(M_Z)=0.119.$ Figure~\ref{fig:chi2-vs-alps-pseudo} shows this idealized distribution rendering a deeper parabola.

The flatter $\chi^2_R$ of the clusters of actual residuals in Fig.~\ref{fig:chi2-vs-alps-K-folds} indicates 
a larger uncertainty than $\approx 0.0003$ obtained by fitting the ideal parabola in Fig.~\ref{fig:chi2-vs-alps-pseudo}, with the latter being in a perfect agreement with $\Delta \chi^2=1$. The second part of this exercise tries to estimate the actual uncertainty, keeping in mind that it depends on how the clusters are formed. To gain an insight about the practical K-folding studies, we determine $\delta\alpha_s$ from random combinations of three out of four above clusters. In this specific example, and using the DT, we find that $\delta\alpha_s$ varies between $\approx$ 0.0002 and 0.002 depending on the magnitude of tension in the dominant cluster of each configuration. 

This exercise can be expanded in a more realistic cross-validation setup to explore the variability of $\alpha_s(M_Z)$ and its uncertainty across the full granularity of data. Appendix~\ref{app:ToyKFolding} contains further details, including concrete uncertainty estimates.


\section{Bayesian Models for Uncertainty Estimation }
\label{sec:BayesianModels}

The fundamental assumption of the tolerance prescriptions is that all
experiments agree on the parameter of interest in a certain region,
and this region can be estimated from the augmented loglikelihood by
following the described procedures. In contrast, Bayesian model
averaging (BMA) assumes that some experiments are more likely to be
correct than others. BMA estimates the underlying probability for each
experiment to be correct according to the data. In this section, we describe how two such models can be used to determine the best-fit $\alpha_s$ and its uncertainty, working with the CT25 baseline data set as an example. We focus on combination of $\alpha_s$ values preferred by DIS, DY, and Jet+$t\bar t$ classes of processes summarized in Table~\ref{tab:gaussian_combination}. As in the previous section, we make a simplifying assumption that the correlations among the separate experimental data sets via the PDFs or other parameters are negligible, allowing us to treat these data as independent $\alpha_s$ determinations.

\subsection{A Bayesian Hierarchical Model}
\label{sec:BHM}
In combining the different values of $\alpha_{s,i}(M_Z)$ from multiple experiments, a key difficulty is that the quoted uncertainties from each experiment may not fully capture all sources of systematic error, theory assumptions, or hidden correlations.
A naive combination, such as the simple weighted average in Sec.~\ref{sec:PDG-Uncertainty-combination}, implicitly assumes that all measurements are mutually consistent and that their reported uncertainties are complete. When this assumption fails, the resulting average can become artificially precise or biased toward experiments that happen to quote smaller, but possibly underestimated, error bars.

A Bayesian hierarchical model (BHM) provides a principled way to address this problem. Instead of treating each experimental determination $\alpha_{s_i}(M_Z)$ as an estimate of a single common parameter with known variance, the hierarchical approach acknowledges the possibility of unmodeled spread among the measurements by introducing hyperparameters that describe a latent ``parent'' distribution from which the individual measurements of $\alpha_{s,i}(M_Z)$ are drawn. 
Specifically, considering each $\alpha_{s_i}(M_Z)$ to be the result of a random draw from a parent distribution, one can write
\begin{equation}
\label{eq:hyperdraw}
p(\alpha_{s_i})=\int p(\alpha_{s_i}|\mu,\tau)p(\mu,\tau)d\mu d\tau\ ,
\end{equation}
where $p(\mu,\tau)$ is a hyperprior distribution for the hyperparameters $\mu$ and $\tau$.

Before introducing the specific hierarchical model used in this analysis, it is important to comment on the choice of hyperprior distribution. In Bayesian hierarchical methods, the role of the hyperprior is not to encode detailed knowledge about the data, but rather to parameterize our uncertainty about the possible degree of inter-experimental variation.
While a fully data-driven determination of the hyperprior is in principle possible, it generally requires a more complex model structure. We consider two possible models, one where the the choice of hyperprior is well-motivated but not derived from data (in this subsection), and the other where the hyperprior is determined through data-driven methods (in Sec.~\ref{sec:GMM}). 

A suitable guiding principle in choosing the hyperprior is that it should (i) be weakly informative, so that it does not dominate the posterior when the data are mutually consistent, and (ii) be flexible enough to accommodate additional dispersion among the measurements when they exhibit tension. The form proposed in Ref.~\cite{Erler:2020bif} satisfies these requirements while avoiding ad hoc error inflation or outlier removal. It introduces a single hyperparameter $\tau$ describing the possible extra variance beyond the quoted experimental uncertainties, and assigns to it a $\mu$-independent hyperprior, \
\begin{equation}
p(\tau)\, d\tau^2 \propto 
\prod_{i=1}^{N} \left[\frac{1}{\sigma_i^2 + \tau^2}\right]^{\beta/(2N)} d\tau^2,
\label{hyperprior}
\end{equation}
which smoothly interpolates between a flat prior ($\beta = 0$) and a sharply peaked one for larger values of $\beta$. This construction allows the analyst to encode a controlled level of prior belief about the reliability of the quoted uncertainties: smaller values of $\beta$ correspond to a more permissive prior, suitable when the data exhibit significant disagreement, while larger values correspond to the assumption that most of the quoted uncertainties are reliable.
This model was introduced as a more general alternative to the PDG prescription \cite{Erler:2020bif}.

\begin{figure}[ht]
\centering
\includegraphics[width=0.45\textwidth]
{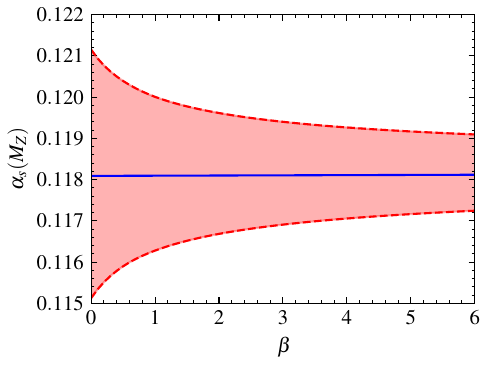}
\includegraphics[width=0.45\textwidth]
{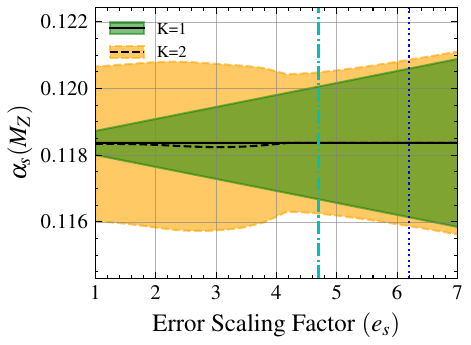}
\caption{ Variation of the uncertainty on $\alpha_s(M_Z)$ when using the BHM (left) with a hyperprior defined in Eq.~\eqref{hyperprior}, and  
 when using the GMM (right) as a function of the error scaling factor $e_S$. The orange shaded region represents the uncertainty when setting $K=2$ in the GMM, and the green shaded region represents the uncertainty for $K=1$, which corresponds to the usual $\chi^2$ fit. Also shown are vertical lines indicating the values of $e_S \approx 4.7$ and $e_S \approx 6.2$, respectively corresponding to the PDG scale factor in Eq.~\eqref{eq:pdg_uncertainty} and tolerance $T$ with standard combination ($K=1$) that would give the same uncertainty estimate as the GMM with $K=2$.
For both plots, the data used in the combination is shown in Table~\ref{tab:gaussian_combination}.
}
\label{fig:alpha_s_combination}
\end{figure}

We combine the data given in Table~\ref{tab:gaussian_combination}, according to which $N=3$ in Eq.~(\ref{hyperprior}) represents the number of $\alpha_s(M_Z)$ measurements being combined, and $\sigma_i$ are the corresponding uncertainties. For $\beta =0$, we find $\alpha_s(M_Z) = 0.1181^{ + 0.0030}_ {- 0.0031}$.
The variation of the uncertainty estimate as a function of $\beta$ is shown in Fig.~\ref{fig:alpha_s_combination} (left), where we note that the $\alpha_s(M_Z)$ uncertainty reduces as $\beta$ increases. Specifically, for $\beta =15$ we find $\alpha_s(M_Z) = 0.1182^{ + 0.0006}_ {- 0.0007}$. 
Given the large spread of $\alpha_s(M_Z)$ values in
Table~\ref{tab:gaussian_combination}, and following the suggestion of
Ref.~\cite{Erler:2020bif}, we use smaller values of $\beta$. Rather
than choosing a single value of $\beta$, we average over
$0\le\beta\le1$,  which yields
\begin{equation}
\alpha_s(M_Z) = 0.1181^{ + 0.0022}_ {- 0.0023} \ .
\label{eq:bhm_uncertainty}
\end{equation}
%

\subsection{A Gaussian Mixture Model}
\label{sec:GMM}

The Gaussian mixture model (GMM) proposed in Ref.~\cite{Yan:2024yir} is a new Bayesian hierarchical model in which the hyperprior is determined from data through the use of statistical information criteria. In this section, we summarize an application of the GMM to combine the $\alpha_s(M_Z)$ values with a large spread in Table~\ref{tab:gaussian_combination} by capturing the multimodal feature of these data. Appendix~\ref{app:GMM_AIC} discusses technical aspects of this combination.

In the GMM, $K$ Gaussian components are introduced, so that the likelihood is modeled as
\begin{equation}
L(\alpha_s) = \sum_{k=1}^K w_k \, \mathcal{N}(\alpha_s \mid \mu_k, \sigma_k^2),
\label{eq:gmm_likelihood}
\end{equation}
where $w_k$ are mixture weights. The optimal value of $K$ is determined using model selection criteria such as the Akaike Information Criteria (AIC)~\cite{Akaike} and Bayesian Information Criteria (BIC)~\cite{BIC}. For the present combination, the choice of $K = 2$ minimizes both the AIC and BIC. Note that $K = 1$ corresponds to the standard $\chi^2$ combination in Table~\ref{tab:gaussian_combination}. Minimizing the AIC score with respect to $K$ yields a bimodal posterior for $\alpha_s(M_Z)$. Following the procedure of Ref.~\cite{Yan:2024yir}, we extract the $\alpha_s(M_Z)$ central value and its uncertainty from this distribution and find
\begin{align}
\alpha_s &= 0.1183 \pm 0.0023 \quad (\text{GMM, } K=2)\ .
\label{eq:GMM_uncertainty}
\end{align}

Importantly, this estimate is obtained without inflating the errors, i.e., for the error rescaling factor $e_S=1$. We have also examined the combination under the tentative scaling of errors by a factor $e_S>1$. We find that the $K=2$ combination with $e_S\approx 1$ is always preferred (has the lowest AIC score), closely followed by the traditional $K=1$ combination with $e_S \approx 3.5$ that has a marginally higher score -- see Fig.~\ref{fig:aic_vs_es} in Appendix~\ref{app:GMM_AIC}.  Thus, it would not be unreasonable to also consider the $K=1$ case when $e_S\approx 3.5$ to determine the $\alpha_s$ uncertainty. This choice, however,  produces a poor agreement with the input data in Table~\ref{tab:gaussian_combination}, with a large value of $\sum_i\chi^2(\overline\alpha_s,\alpha_{s,i},\sigma_i)$ as well as a too small uncertainty. We therefore do not pursue it.   

Figure~\ref{fig:alpha_s_combination} (right) illustrates the dependence of the $\alpha_s(M_Z)$ uncertainty on $e_S$ for both $K = 1$ (standard $\chi^2$) and $K = 2$ (GMM). The orange and green shaded regions represent the uncertainty estimates for $K=2$ and $1$, respectively. When the errors are not rescaled, then $e_S = 1$, and $K=1$ produces a very small uncertainty identical to the one presented in the last column of Table~\ref{tab:gaussian_combination}. On the other hand, for this $e_S$ value (specifically, for $e_S < 2.3$), the $K=2$ model is preferred according to the AIC and BIC criteria. The uncertainty for $K=1$ increases with $e_S$, whereas the uncertainty for $K=2$ decreases around $e_S \approx 4.2$ and then increases linearily as in the $K=1$ model, see Figure~\ref{fig:alpha_s_combination}. 

From Fig.~\ref{fig:alpha_s_combination} (right), it follows that the standard ($K=1$) treatment reproduces the $\alpha_s$ uncertainty obtained with the GMM ($K=2$) by applying  $e_S\approx 6.2$, or, equivalently, adopting a tolerance $T$ such that $\Delta\chi^2 = T^2 \approx 33$ (see Sec.~\ref{sec:PDG-Uncertainty-combination} for the relation between $T$ and $e_S$):
\begin{align*}
    \alpha_s &= 0.1183 \pm 0.0023 \quad (\text{scaled errors, } e_S \approx 6.2 \text{ or } T^2 \approx 33).
\end{align*}
This is larger than $e_S=4.7$ from the PDG prescription in Sec.~\ref{sec:PDG-Uncertainty-combination} and slightly smaller than the effective scaling of $e_S\approx 6.9$, or $T^2\approx 48$, that reproduces the uncertainty estimate from the  global tolerance criterion defined in Sec.~\ref{sec:Global-Dynam}. In Fig.~\ref{fig:alpha_s_combination} (right), the values of $e_S=4.7$ and 6.2 are indicated by vertical dot-dashed and dotted lines.


\section{Final combination of \texorpdfstring{$\alpha_s$}{QCD coupling} determinations}
\label{sec:combination}
Having reviewed the physics factors affecting the extraction of $\alpha_s$ and several procedures to determine the $\alpha_s$ uncertainty, we are ready to combine this information to provide the final $\alpha_s$ determination.

In Sec.~\ref{sec:DeltaChi2Eq1}, we pointed out that the $\Delta\chi^2 = 1$ criterion would capture only a small part of realistic $\alpha_s$ uncertainty, and that an additional uncertainty must be associated with data's internal consistency.
As a reminder, when examining the $\chi^2$ profile for the CT25prel baseline Fig.~\ref{fig:alpha_s_chi2_scan}, the application of the $\Delta \chi^2 = 1$ criterion to the total $\chi^2$ (black curve) would result in $\alpha_s (M_Z) = 0.11837^{+ 0.00035}_{- 0.00041}$. This uncertainty is smaller than the global average quoted in the PDG~\cite{ParticleDataGroup:2024cfk}. 

As evidence supporting this conclusion, Sec.~\ref{sec:results-P2} and Sec.~\ref{sec:PDG-Uncertainty-combination} pointed out a significant spread in the minima of $\chi^2$ profiles from different classes of experiments (DIS, DY, and Jet and top-quark pair
production (Jet+$t{\bar t}$) combined), reflecting the presence of tension between the experiments that has been also confirmed at the level of individual experiments by other techniques. We also noted the sensitivity of the $\chi^2$ profiles to only partly known normalizations of the correlated systematic errors controlled by the \texttt{ksys} parameter. In particular, we noticed that the DIS measurements show significant disagreement with DY and Jet+$t\bar{t}$ measurements for the default normalizations of systematic errors. 

These limitations of the $\Delta\chi^2 = 1$ criterion raise the question of finding more complete procedures to quantify the uncertainty. We have explored several such procedures in Sec.~\ref{sec:Tolerances} and Sec.~\ref{sec:BayesianModels}. In this section we compile the results from these different prescriptions and combine them into one final estimate of the $\alpha_s$ central value and uncertainty.

\subsection{The central \texorpdfstring{$\alpha_s$}{QCD coupling} value}
In Sec.~\ref{sec:pdfs}, we observed that the choice between the acceptable baseline data sets affects the central $\alpha_s$ value. The two described baselines, CT25prel and CT25, yield the central values of about 0.1179 and 0.1184, respectively. In the following, we will quote the latter as our final $\alpha_s$ determination. In contrast, the choice between the baselines only weakly influences the uncertainty determination, which instead depends on the method for the estimation. In the following, we summarize the uncertainty ranges obtained with several such methods.

\subsection{Final  \texorpdfstring{$\alpha_s$}{QCD coupling}  determination}
\label{sec:OtherDeterminations}
The four approaches that we choose for the final combination are based on the global and dynamical tolerances, as well as on the two Bayesian models.

{\bf Global and dynamical tolerances.} Sec.~\ref{sec:Global-Dynam} described the commonly used uncertainty prescriptions based on the global and dynamical tolerance, or GT and DT. When we apply the GT criterion based on the 68\% quantile for the sum of all $\chi^2_{E}$ using  the CT25 baseline from the individual experiments, we obtain
\begin{equation}
\delta \alpha_s(M_Z) = {0.1184}^{+ 0.0026}_{ - 0.0028} \ .   
\label{eq:global_tolerance}
\end{equation}
The DT procedure results in 
\begin{equation}
\delta \alpha_s(M_Z) = {0.1184}^{ + 0.0023}_ {- 0.0014}, 
\label{eq:dynamical_tolerance}
\end{equation}
which exhibits substantial asymmetric uncertainty.

These results are obtained for the default pole mass $m_c=1.3$ GeV of the CT25 analysis. While in principle the charm mass is correlated with the gluon PDF and therefore $\alpha_s$, in practice this correlation is weak. Appendix~\ref{app: mc-var} shows the GT and DT $\alpha_s$ ranges for the alternative masses 1.2 and 1.4 GeV, and those essentially coincide with the estimates for 1.3 GeV. The $m_c$ dependence is hence found to be minimal, as is also illustrated by the $\chi^2$ profiles in Fig.~\ref{chi2-vs-alps}.   

{\bf Bayesian Models}. In Section~\ref{sec:BayesianModels}, we discussed two Bayesian models to estimate $\alpha_s$ uncertainties. 
We obtained $\alpha_s(M_Z) = 0.1181^{ + 0.0022}_ {- 0.0023}$
with the Bayesian Hierarchical Model and $\alpha_s(M_Z)=0.1183 \pm 0.0023 $ with the Gaussian Mixture Model.

{\bf The average of four methods.}
\begin{table}[]
    \centering
    \begin{tabular}{|c|c|c|}
    \hline
        Statistical Method & Eq. & $\delta\alpha_s(M_Z)$ \\\hline \hline
       Global Tolerance& \ref{eq:global_tolerance} 
       & ${0.1184}^{+ 0.0026}_{ - 0.0028}$ \\
       Dynamical Tolerance& \ref{eq:dynamical_tolerance} & ${0.1184}^{+ 0.0024}_{ - 0.0012}$\\
       BHM& \ref{eq:bhm_uncertainty} 
       & $0.1181^{ + 0.0022}_ {- 0.0023}$\\
       GMM& \ref{eq:GMM_uncertainty}
       & $0.1183^{+0.0023}_{-0.0023}$ \\ \hline 
       Average
       & & ${0.1183}^{ + 0.0023}_{ - 0.0020}$ \\ \hline
    \end{tabular}
    \caption{Estimates of $\alpha_s(M_Z)$ obtained by four distinct analyses of statistical uncertainty. }
    \label{tab:final_average}
\end{table}

Table~\ref{tab:final_average} collects the uncertainty estimates obtained with our four methods. The last row, labeled ``Average", corresponds to approximating the $\alpha_s(M_Z)$ probability distribution from each statistical model as a split normal distribution and summing over each model with uniform weights to produce an average. This yields a final combined average value and uncertainty of $\alpha_s(M_Z) = {0.1183}^{+0.0023}_ {-0.0020}$ .   

Good consistency among the four determinations increases our confidence in this result. It is important to emphasize that the determinations do not invoke a fixed tolerance value. Instead, they would justify a specific $T^2$, should it be used as a proxy for the more complex approaches that we describe. For the CT25 baseline, the average $\delta\alpha_s(M_Z)$ in Table~\ref{tab:final_average} translates into the $\Delta\chi^2={}^{+39}_{-25}$ for the global $\chi^2$. This is close, although not identical, to the approximate $T^2=37$ at 68\% CL introduced in the ``CT two-tier tolerance'' prescription used in the previous CT PDF analyses~\cite{Gao:2013xoa,Dulat:2015mca,Hou:2019efy}. 
\section{Conclusions}
\label{sec:conclusion}

We have presented a new determination of the strong coupling constant
$\alpha_s(M_Z)$ from the CT25 global QCD analysis at NNLO, incorporating
high-precision LHC Run-2 measurements and improved statistical
methodology. The global analysis simultaneously fits parton distribution
functions, $\alpha_s$, and other QCD parameters to a diverse collection of
measurements from fixed-target and collider experiments, providing the
most complete treatment of correlations among these parameters. Our
central result, combining four independent uncertainty prescriptions, is
\[
\alpha_s(M_Z) = 0.1183^{+0.0023}_{-0.0020} \quad (68\%~\mathrm{CL}).
\]

This value is in good agreement with the current PDG world average and with
other recent determinations from LHC data and lattice QCD.

A central theme of this work is the critical assessment of uncertainty
quantification. We demonstrated that the naive $\Delta\chi^2 = 1$
uncertainty—approximately $\pm 0.0004$ for the CT25 data set—is
insufficient, because it captures only statistical fluctuations and
ignores the significant tensions among the three broad classes of data
(DIS, Drell--Yan, and jet~+~$t\bar{t}$ production). These tensions are
reflected in the large spread of preferred $\alpha_s$ values across
process types (0.1152 to 0.1198 for the CT25prel baseline with
multiplicative systematic errors) and in the failure of the combination
of the three classes to satisfy a standard goodness-of-fit test by a
wide margin.

We investigated four methodologically distinct approaches to incorporate
the full uncertainty:
\begin{itemize}
	
	\item \textbf{Global tolerance (GT)}: Based on the 68\% quantile of the
	total $\chi^2$ distribution, yielding
	\[
	\alpha_s(M_Z) = 0.1184^{+0.0026}_{-0.0028}.
	\]
	It is trivially insensitive to data clustering.
	
	\item \textbf{Dynamical tolerance (DT)}: Constructed from quantiles of the
	individual experiment profiles, giving
	\[
	\alpha_s(M_Z) = 0.1184^{+0.0024}_{-0.0012},
	\]
	with a pronounced asymmetry. We showed that DT estimates depend on how
	experiments are grouped into clusters, and we introduced the concept of
	\emph{data-clustering safety} to characterize this dependence: an
	uncertainty estimate on a fundamental parameter is clustering-safe if it
	remains stable under reasonable regroupings of the data. The GT satisfies
	this criterion by construction; the DT does not in general and requires
	supplementary tests. In the limit of a single cluster the DT reduces to
	the GT, while in the limit of many clusters it approaches the
	$\Delta\chi^2=1$ uncertainty.
	
	\item \textbf{Bayesian hierarchical model (BHM)}:
	Introducing a hyperprior that accounts for unmodeled inter-experimental
	spread, and averaging over the hyperprior parameter $0 \le \beta \le 1$,
	yields
	\[
	\alpha_s(M_Z) = 0.1181^{+0.0022}_{-0.0023}.
	\]
	
	\item \textbf{Gaussian mixture model (GMM)}:
	Determining the data-adaptive hyperprior via information criteria, we find a bimodal posterior driven by multimodal structure in the data which yields
	\[
	\alpha_s(M_Z) = 0.1183 \pm 0.0023.
	\]
	
\end{itemize}

The four methods are in good mutual agreement. Their uniformly weighted
combination gives the final value quoted above, corresponding to an
effective global tolerance of $\Delta\chi^2 = {}^{+39}_{-25}$, close to
the $T^2 = 37$ used in previous CT PDF analyses.

We also identified systematic uncertainties that could shift the central
value. The treatment of correlated systematic errors in inclusive-jet
production is the most consequential: switching the normalization
convention from multiplicative (\texttt{ksys=1}) to additive
(\texttt{k=2}) for jet data alone lowers the global $\alpha_s$ by
about 0.003, while simultaneously producing a large internal tension
within the jet data sets and a substantially poorer goodness-of-fit.
Since the major jet-energy-scale uncertainties at the LHC are regarded as
multiplicative, the \texttt{ksys=1} result is preferred; nonetheless, this
exercise underscores the importance of clear documentation of
systematic-error conventions by experimental collaborations. We also
found that the choice of PDF parametrization and variations of the
charm-quark pole mass between 1.2 and 1.4~GeV have only a small effect on
$\alpha_s$, with the parametrization uncertainty confined to
$\approx \pm 0.0005$ across 287 independent candidate parametrizations.

Looking ahead, future HL-LHC measurements will further constrain
$\alpha_s$ and the PDFs, but the precision gains will be meaningful only
if accompanied by robust uncertainty quantification. The concept of
data-clustering safety introduced here provides a concrete, calculable
diagnostic for the stability of uncertainty estimates based on tolerance
or cross-validation methods, applicable to both Hessian and Monte Carlo
PDF methodologies. Together with the Bayesian approaches discussed in this work, the proposed methodologies provide a framework for self‑consistent uncertainty quantification, which is  essential for a precise determination of $\alpha_s$ from hadron‑collider data.

\section*{Acknowledgments}
We thank Amanda Cooper-Sarkar, Olaf Behnke, Tom Cridge, Louis Lyons, Robert Thorne, CTEQ-TEA colleagues, and participants of PDF4LHC-PHYSTAT meetings for conversations on uncertainty quantification for QCD parameters determined in PDF fits. We thank Tie-Jiun Hou for help on the CT25 analysis. The work of MG is partially supported by the National Science Foundation under Grant No.~PHY-2412071. KM was supported in part by the US National Science Foundation under Grant PHY-2310497. PMN and SD are grateful for support from the Wu-Ki Tung Endowed Chair in particle physics. The work of SD was also supported by the National Natural Science Foundation of China under Grant number 12475079. The work of CPY is supported by the U.S. National Science Foundation under Grant No. PHY-2310291.

\appendix

\section{New LHC data sets included in the CT25 NNLO analysis
\label{app:NewData}}
In this appendix, we provide additional details on the newly included data sets listed in Table~\ref{Tab:LHCnewData} of the main narrative.

{\bf Lepton pair production measurements}. As in the CT18A global fit, we include the precision measurement of $W$ and $Z$ production by ATLAS collaboration at 7 TeV \cite{Aaboud:2016btc}. This measurement revealed an enhancement in the strangeness PDF at $x\sim 0.02$ -- a trend that was corroborated by more recent observations at the LHC. In Ref.~\cite{Sitiwaldi:2023jjp}, we examined in depth the implications of these measurements for PDFs and identified the following differential cross sections for inclusion in the final CT25 fit:
\begin{enumerate}
\item From the ATLAS collaboration: 
\begin{enumerate}
\item muon pseudorapidity distribution in $W$ boson production $(W\rightarrow \mu \nu)$ at a center-of-mass energy $\sqrt{s} = 8$ TeV~\cite{ATLAS:2019fgb} with 20.2 fb$^{-1}$ of integrated luminosity (IL), in fiducial volume $p^{l,\nu}_T > 25$ GeV, $|\eta_{\mu}| < 2.4$. $m_T > 40$ GeV; 
\item $W^{\pm}$ muon pseudorapidity and $Z$ dilepton rapidity distributions~\cite{ATLAS:2018pyl} in $\sqrt{s}=5.02$ TeV with 25 pb$^{-1}$ of IL, in the fiducial phase space $p^{l,\nu}_T > 25$ GeV, $|\eta_{l}| <2.5$, $m_T >40$ GeV for $W^{\pm}$ bosons and $p^{l}_T > 20$ GeV, $|\eta_{l}| <2.5$, $66< m_{ll} <116$ GeV for $Z$ bosons;
\item triple differential distribution in $Z$ boson production in electron and muon decay channels \cite{ATLAS:2017rue}, presented in terms of invariant mass $m_{ll}$, rapidity of dilepton pair $y_{ll}$, and cosine of the polar angle in the Collins-Soper frame, with an IL of 35.9 fb$^{-1}$, $p^{l}_T > 20$ GeV, $|\eta_{l}| <2.4$, $46< m_{ll} <200$ GeV.
\end{enumerate}

\item From the CMS collaboration: 

rapidity distribution for $Z$ boson production at 13 TeV~\cite{CMS:2019raw} with an IL of 35.9 fb$^{-1}$,  $p^{l}_T > 20$ GeV, $|\eta_{l}| <2.4$ and  $|m_{ll}-M_Z| < 15$ GeV.

\item From the LHCb collaboration: 
\begin{enumerate}
\item electron pseudorapidity differential cross section in $W \rightarrow e\nu$ production ~\cite{LHCb:2016zpq} at $\sqrt{8}$ TeV with an IL of 2 fb$^{-1}$,  $2.0 < \eta < 4.25$ and $p^e_T > 20$ GeV;
\item  dilepton rapidity distribution in $Z$ boson production in electron and muon decay channels at $\sqrt{s}=13$ TeV~\cite{LHCb:2021huf} with 0.29 fb$^{-1}$ of IL, $2.0 < \eta_l < 4.5$, $p^l_T > 20$ GeV, $60~<~m_{ll}~<~120$ GeV. 
\end{enumerate}
\end{enumerate}

{\bf Top-quark pair production measurements}. For top-quark pair production, we included the following measurements 
at $\sqrt{s} = $ 13 TeV based on the study of their impact on post-CT18 PDFs in Ref.~\cite{Ablat:2023tiy}:
\begin{enumerate}
    \item From the ATLAS collaboration: 
\begin{enumerate}
\item rapidity distribution of the $t\bar t$ pair, $d\sigma/dy_{t\bar t}$, in the all-hadronic channel~\cite{ATLAS:2020ccu} with an IL of 36.1 fb$^{-1}$ in the full phase space at parton level; 
\item the statistically combined $m_{t\bar t}$ + $y_{t\bar t}$ + $y^B_{t\bar t}$ + $H^{t\bar t}_T$ distributions~\cite{ATLAS:2019hxz}, with 36 fb$^{-1}$ of IL, where $y^B_{t\bar t}$ denotes the reconstructed rapidity of the $t \bar t$ system in the boosted topology, and $H^{t\bar t}_T$ is the scalar sum of the transverse momenta of the hadronic and leptonic top quarks, $H^{t\bar t}_T = p^{had}_{t} +p^{lep}_{\bar t}$.
\end{enumerate}
\item From the CMS collaboration: 
\begin{enumerate}
\item the $y_{t\bar t}$ distribution in the dilepton channel~\cite{CMS:2018adi} with
35.9 fb$^{-1}$ of IL;
\item the $m_{t\bar t}$ distribution in the  lepton+jet channel~\cite{CMS:2021vhb} with 137 fb$^{-1}$ of IL.
\end{enumerate}
\end{enumerate}

{\bf Single-inclusive jet production measurements}. 
In Ref.~\cite{Ablat:2024uvg}, we compared constraints imposed on our post-CT18 PDFs by the latest LHC cross sections in hadronic jet production. While these cross sections can be included in the form of either inclusive single-jet or jet-pair distributions, Ref.~\cite{Ablat:2024uvg} found it preferable to fit the single-inclusive jet cross sections only, in light of the stronger dependence of theory predictions in dijet production on the choice of renormalization and factorization scales. Based on the detailed examination of candidate jet cross sections in Ref.~\cite{Ablat:2024uvg}, the CT25 analysis includes the following single-inclusive jet production data sets:
\begin{enumerate}
\item From the ATLAS collaboration:
\begin{enumerate}
\item the $d^2\sigma/(dp_Td|y|)$ distribution at $\sqrt{s} = 8$ TeV~\cite{ATLAS:2017kux} with 20.2 fb$^{-1}$ of IL, 
jet radius $R={0.6}$, 
in the fiducial volume $|y|<3$ and $70\leq p^{jet}_T\leq 2500$ GeV; 
\item the $d^2\sigma/(dp_Td|y|)$ distribution at $\sqrt{s} = 13$ TeV~\cite{ATLAS:2017ble} with 3.2 fb$^{-1}$ of IL,  $R=0.4$, 
$|y|<3$ and $100\leq p^{jet}_T\leq 3937$ GeV.
\end{enumerate}

\item From the CMS collaboration, 

\begin{enumerate}

\item
In the CT25prel fit, we include $d^2\sigma/dp_Td|y|$ distributions at $\sqrt{s} = 13$ TeV~\cite{CMS:2016lna} with 33.5 fb$^{-1}$ of IL, $R={0.7}$, $|y|<2$ and $97\leq p^{jet}_T\leq 3103$ GeV. 

\item

In the CT25 fit, we include $d^2\sigma /(dp_T dy)$ distributions at $\sqrt{s}=7~\mathrm{TeV}$ with an IL of $5~\mathrm{fb}^{-1}$ (ID=556)~\cite{CMS:2012ftr,CMS:2014nvq,CMS:2014qtp,CMS:2024trs}, at $\sqrt{s}=8~\mathrm{TeV}$ with an IL of $19.7~\mathrm{fb}^{-1}$  (ID=557)~\cite{CMS:2016lna,CMS:2024trs}, and at $\sqrt{s}=13~\mathrm{TeV}$ with an IL of 33.5 fb$^{-1}$  (ID=555)~\cite{CMS:2021yzl,CMS:2024trs}. The $7~\mathrm{TeV}$ measurements include 118 data points in six rapidity bins covering $0 \leq |y| \leq 2.5$, with jet transverse momenta in the range $114 \leq p_T \leq 1327~\mathrm{GeV}$. The $8~\mathrm{TeV}$ measurements include 164 data points in six rapidity bins covering $0 \leq |y| \leq 3.0$, with $74 \leq p_T \leq 1784~\mathrm{GeV}$. The $13~\mathrm{TeV}$ measurements include 78 data points, $|y|<2$ and $97\leq p^{jet}_T\leq 3103$ GeV. 
These CMS 7, 8, 13 TeV measurements have updated systematic uncertainties and newly provided statistical correlation matrices~\cite{CMS:2014qtp,CMS:2024trs}. They supersede the earlier inclusive-jet datasets at $7~\mathrm{TeV}$ (ID=542)~\cite{CMS:2012ftr,CMS:2014nvq},  $8~\mathrm{TeV}$ (ID=545) \cite{CMS:2016lna}, and $13~\mathrm{TeV}$ (ID=555)~\cite{CMS:2021yzl} included in previous analyses~\cite{Hou:2019efy,Ablat:2024uvg}.
\end{enumerate}
\end{enumerate}

\section{Investigation of reclustering prescriptions}
\label{app:Reclustering}
\begin{figure}[p]
\centering
\includegraphics[width=0.8\linewidth]{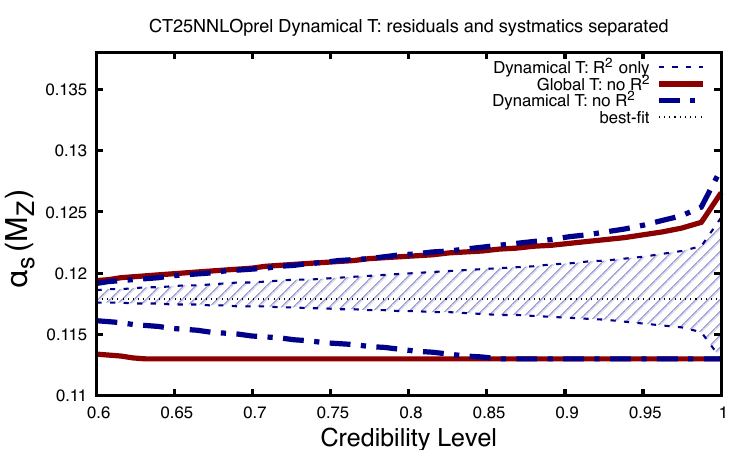}
\caption{Contributions to the GT and DT uncertainties vs. the credibility level for the CT25prel baseline from nuisance parameters ($R^2$) only and from residuals ($D^2$) without $R^2$.}
\label{fig:alps-vs-CredLevSeparate}
\end{figure}
\begin{figure}[p]
\centering
\includegraphics[width=0.8\linewidth]{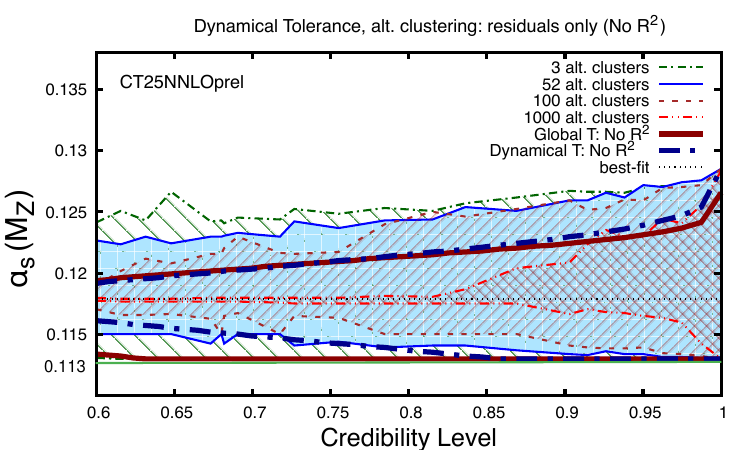}
\caption{Reclustering without $R^2$: $\alpha_s{(M_Z)}$ as a function of the credibility level for the dynamical and global tolerance criteria compared to different alternative cluster configurations.}
\label{fig:alps-vs-CredLev-alt-clust}
\end{figure}

In continuation of the discussion in Sec.~\ref{sec:reclustering_algorithm}, we illustrate the possibility of separate reclustering of the residuals and nuisance parameters by two figures. Figure~\ref{fig:alps-vs-CredLevSeparate} goes back to the example of the GT and DT uncertainties for the published data sets in Fig.~\ref{fig:alps-vs-CredLev} for the CT25 baseline and now shows the DT uncertainties constructed only with $D^2$ and only with $R^2$ for the CT25prel baseline.\footnote{These uncertainties do not change much between the two baselines. In this and following figure, the lower bands saturate at $\alpha_s(M_Z)=0.113$, the lowest point on our $\alpha_s$ grid.} The GT and DT uncertainties based only on residuals (i.e., without $R^2$) are wider than the corresponding ones based on both residuals and nuisance parameters ($D^2+R^2$). 
The role of $R^2$ appears to be somewhat different for the GT and DT cases. In the DT case, the $R^2$ has a more constraining power on the uncertainty:  if it were to be the only metric for the uncertainty, we would obtain a very small error (blue dashed band).  However, $R^2$ is in general not perfectly determined, due to lack of full control on the correlated systematic uncertainties. The $R^2$ effect is less pronounced for the GT. 

Figure~\ref{fig:unc-summary} showed how the 68\% and 90\% CL error estimates change when we vary the DT clustering procesure. Figure~\ref{fig:alps-vs-CredLev-alt-clust} extends this comparison to show error estimates for other credibility levels, including the residuals only (that is, neglecting $R^2$). In this figure, the ``Global T: no $R^2$'' and ``Dynamical T: no $R^2$'' refer to the GT and DT uncertainties based on the full baseline. 
The ``52 alt. clusters'' and other lines refer to randomly generated clusters constructed by reshuffling the residuals in the 52 experimental data sets of CT25prel. For the alternative clusters, the lines show the envelopes of the DT estimates for 200 shuffles of the corresponding alternative clusters.  
In an ideal scenario, in the absence of tensions among the data, alternative cluster configurations would produce approximately the same estimate for the $\alpha_s(M_Z)$ uncertainty. In reality, when we partition into a lot of clusters, taking the outcome of the DT without $R^2$ by itself may already produce a stringent (likely, too stringent) uncertainty on $\alpha_s(M_Z)$. However, the constraints may be weaker for other cluster configurations, and including $R^2$ into clustering may additionally reduce or increase the strength of the DT constraints by modifying the tensions, similarly to what we saw for the DT that reclustered the published data sets.

\section{A toy example of uncertainty under K-folding}
\label{app:ToyKFolding}

\begin{table}[b]
\centering
\begin{tabular}{|ccccc|}
\hline
 & cluster 0\quad\quad & cluster 1\quad\quad & cluster 2\quad\quad & cluster 3 \\
\hline
$N_{\textrm{pt}}$ & 1113 & 1113 & 1112 & 1112 \\
\hline
  \multicolumn{5}{|c|}{\quad\quad\quad\quad $\chi^2_R$ (no $R^2$)} \\ 
  configuration 1 & 1094 & 1157 & 1281 & 1177 \\ 
  configuration 2 & 1102 & 1158 & 1252 & 1198 \\ 
  configuration 3 & 1065 & 1239 & 1251 & 1156 \\
\hline
\end{tabular}
\caption{Numbers of points and $\chi^2_R$ (without the systematic contribution $R^2$) for three random partitions (configurations) of the residuals into four pseudodata clusters.}
\label{tab:chi2-folds}
\end{table}

This appendix provides details about the toy study of K-folding summarized in Sec.~\ref{sec:K-folding}.
Working with the best-fit residuals and nuisance parameters (assumed to be mutually independent) in the setup of Sec.~\ref{sec:reclustering}, we divide the baseline data into four clusters according to randomly generated partitions (``configurations''). In contrast to the full K-folding procedure, we will not refit (optimize) the PDFs for each configuration, which would be too computationally expensive. Instead, we examine uncertainties by reclustering the output of the CT25prel global fit, similarly to what was done in the previous sections. In addition, to assess their independent impact, we will separately consider the residual and nuisance-parameter contributions. 

{\bf Dependence of the central $\alpha_s(M_Z)$ on clustering.}
We generate multiple partitions of the residuals from $\mathbf{v}_r$ into four clusters, with about 1113 points in each cluster, as outlined in Sec.~\ref{sec:reclustering_algorithm}.
Table~\ref{tab:chi2-folds} reports the numbers of points and $\chi^2$ contributions for three representative configurations in our series. Similarly, we construct four ``$R^2$'' clusters consisting of the shuffled best-fit nuisance parameters from $\mathbf{v}_\lambda$.
Figure~\ref{fig:chi2-vs-alps-K-folds} illustrates the outcome of one such exercise, in which 100 configurations of four clusters were generated. The figure separately depicts the individual $\chi^2/N_{pt}$ ratios for the residuals (purple lines) and nuisance parameters (red-dashed lines). Each purple (red) line corresponds to a different configuration. The black lines represent the $\chi^2/N_{pt}$ average values. The stars represent the minima of the average curves which indicate the $\alpha_s(M_Z)$ central-value preference. No tolerance criterion is applied here; we simply calculate the $\chi^2/N_{pt}$ for each configuration at different $\alpha_s(M_Z)$.

One goal of this exercise is to determine whether the random fluctuations due to the reshuffling process are compatible with their ${\cal N}(0,1)$ distribution expected when the data sets mutually agree. The $\chi^2/N_\textrm{pt}$ vs. $\alpha_s(M_Z)$ for the latter case is shown in Fig.~\ref{fig:chi2-vs-alps-pseudo}, for which we generated an ensemble of the ``cluster 0'' configurations from pseudoresiduals that were generated according to ${\cal N}(0,1)$ from a baseline pseudodata without tensions and assuming the ``truth'' $\alpha_s(M_Z)=0.119$. 

It is therefore interesting to confront the idealized pseudoresidual scenario against the realistic distributions of the residual and $R^2$ clusters. In the ideal case in Fig.~\ref{fig:chi2-vs-alps-pseudo}, the clusters follow a deep parabola centered at the ``truth'' $\alpha_s(M_Z)$. In the real case  in Fig.~\ref{fig:chi2-vs-alps-K-folds}, all clusters have shallower dependence on $\alpha_s(M_Z)$. The parabolic behavior is particularly wide for the residuals and slightly narrower for the nuisance parameters. The clusters of the real residuals all prefer an approximately stable value of $\alpha_s(M_Z)\approx 0.1165$. The preferred $\alpha_s(M_Z)$ of the $R^2$ clusters varies more, between 0.119 and 0.121 depending on the cluster. We read off these values from the central curves of each cluster, obtained by averaging the $\chi^2/N_\textrm{pt}$ curves over many alternative configurations. The fluctuations around these curves, however, are roughly of the same magnitude as in the idealized case of Fig.~\ref{fig:chi2-vs-alps-pseudo}. These observations suggest that tensions among the data sets primarily flatten the $\chi^2/N_\textrm{pt}$ curves  relatively to the no-tension scenario. They also introduce variability in the preferred values of $\alpha_s(M_Z)$ for the $R^2$ clusters, as different groups of experimental systematic uncertainties  exert substantial and inconsistent pulls on $\alpha_s(M_Z)$.

\begin{table}[tb]
\centering
\renewcommand{\arraystretch}{1.3}
\begin{tabular}{|c|c|c|c|c|}
\multicolumn{5}{c}{configuration 1}  \\
\midrule
$\delta\alpha_s(M_Z)=$ & $\mathbf{{}^{+0.00180}_{-0.00062}}$
 & ${}^{+0.00230}_{-0.00400}$
 & ${}^{+0.00410}_{-0.00490}$
 & ${}^{+0.00830}_{-0.00490}$ \\
in 1-clusters: & \{0\} & \{1\} & \{3\} & \{2\} \\
\& 3-clusters: & \{0,1,2\},  \{0,1,3\}, \{0,2,3\} & \{1,2,3\} & N/A & N/A \\
\midrule
\multicolumn{5}{c}{configuration 2}  \\
\midrule
$\delta\alpha_s(M_Z)=$  & $\mathbf{{}^{+0.00102}_{-0.00038}}$
 & ${}^{+0.00507}_{-0.00490}$
 & ${}^{+0.00515}_{-0.00490}$
 & ${}^{+0.00943}_{-0.00490}$ \\
in 1-clusters: & \{0\} & \{1\} & \{3\} & \{2\} \\
\& 3-clusters: & \{0,1,2\},  \{0,1,3\}, \{0,2,3\} & \{1,2,3\} & N/A & N/A \\
\midrule
\multicolumn{5}{c}{configuration 3}  \\
\midrule
$\delta\alpha_s(M_Z)=$  & $\mathbf{{}^{+0.00016}_{-0.00038}}$
 & ${}^{+0.00480}_{-0.00290}$
 & ${}^{+0.00803}_{-0.00490}$
 & ${}^{+0.00819}_{-0.00490}$ \\
in 1-clusters: & \{0\} & \{3\} & \{1\} & \{2\} \\
\& 3-clusters: & \{0,1,2\},  \{0,1,3\}, \{0,2,3\} & \{1,2,3\} & N/A & N/A \\
\bottomrule
\end{tabular}
\caption{
For each configuration of the K-folding exercise, we report the $\delta\alpha_s$ values found by applying the DT, as well as the one 1-cluster and three 3-clusters for which these $\delta\alpha_s$ are achieved. The columns containing the 
$\delta\alpha_s$ values and clusters are arranged so that the leftmost (rightmost) column contains the lowest (highest) $\delta\alpha_s$.
}
\label{tab:K-folding-res}
\end{table}

{\bf $\alpha_s(M_Z)$ uncertainty in the four-cluster exercise}. 
Figure~\ref{fig:chi2-vs-alps-K-folds} demonstrates that the $\chi^2/N_\textrm{pt}$ dependence of clusters is shallower than in the idealized case for which the $\Delta \chi^2=1$ prescription is applicable. The question arises how to quantify the uncertainty in this situation. As was already stated, in an NNPDF-like approach, the $\alpha_s$ uncertainties from individual iterations would affect the weights for the final $\alpha_s$ averaging, and the final outcome would also reflect refitting the PDFs for each data partition into clusters, as well as averaging over partitions' sizes, which is beyond the scope of this study.

Just as one possibility, if the DT criterion is applied either to
 the ``training set'' consisting of 3 clusters or 1 ``testing'' cluster, it can render the uncertainty that can be substantially larger or smaller than the ${}^{+0.0007}_{-0.0014}$ from the NNPDF study. In its most granular realization, the DT criterion treats each residual as an independent ``experiment''.  Table~\ref{tab:K-folding-res} summarizes the outcome of applying such DT criterion to the three training clusters of the residuals (no $R^2$) and to the one left out. Here we show the DT $\alpha_s$ uncertainties for the same three configurations as in Table~\ref{tab:chi2-folds}. These uncertainties can be either substantially smaller or larger than the NNPDF one, depending on the specific partition. The table also highlights the observation that, within a combination of any three clusters, the DT uncertainty of the whole cluster is equal to the smallest individual DT uncertainty among the constituent clusters.  For this specific example, cluster 0 has the lowest $\chi^2$ for the three configurations, cf. Table~\ref{tab:chi2-folds}. When combined with the other two, this cluster drives the net uncertainty of the combination. For the fourth 3-cluster combination, the uncertainty is set by the second most precise individual cluster. According to the findings in Table~\ref{tab:K-folding-res}, the uncertainty on $\alpha_s(M_Z)$ is not uniformly reproduced by each test cluster, and this is mainly ascribed to the tensions in the data. The results do not change if we consider the average of the errors over a large number of configurations.

%

\section{Information Criteria and the GMM}
\label{app:GMM_AIC}
\begin{figure}
    \centering
    \includegraphics[width=0.5\linewidth]{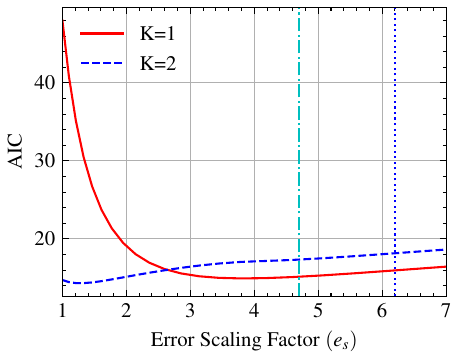}
    \caption{Variation of the AIC score with the error scaling factor $e_S$ shown for both $K=1$ (red) and $K=2$ (blue) in the GMM. Also shown are the values of $e_{S_{PDG}}\approx 4.7$ (dot-dashed) and $e_{S_{GMM}}\approx 6.2$ (dotted) -- see the discussion in Sec.~\ref{sec:GMM}.}
    \label{fig:aic_vs_es}
\end{figure}
We use the Akaike Information Criterion (AIC) to determine the optimal number of Gaussian mixtures $K$ for the GMM.
The AIC score is defined as
\begin{equation}
 \text{AIC} = N_{p} \ \ln N_{\textrm{pt}} - 2 \ln L (\bar{\alpha}_s)
 \label{eq:AIC}
\end{equation}
where $L (\bar{\alpha}_s)$ is the GMM likelihood given in Eq.~(\ref{eq:gmm_likelihood}) evaluated at the optimized value of $\alpha_s$,  $N_{\textrm{pt}}$ is the number of data clusters, and  the number of independent parameters $N_{p}$ for the GMM is given by
\begin{equation}
 N_{p} = K \times N_{\text{par}
 } + ( K - 1 )\ .
 \label{eq:N_par}
\end{equation}
Since we profile the $\chi^2$ over $\alpha_s(M_Z)$, $N_{\text{par} }
=1$. We vary the value of $K$ from $1$ through $5$ and find that the
value of $K$ that minimizes the AIC score is
$K=2$. Our GMM result in Eq.~(\ref{eq:GMM_uncertainty}) quotes the
$\alpha_s$ value for $K=2$.

This result is obtained for the actual input errors corresponding to
error scaling factor $e_S=1$.
In Fig.~\ref{fig:aic_vs_es}, we further illustrate how inflating
the input uncertainties could modify the AIC analysis by plotting the AIC score
vs. $e_S$ for two models: the usual $\sum_i\chi^2(\overline\alpha_s,\alpha_{s,i},\sigma_i)$ in
Sec.~\ref{sec:PDG-Uncertainty-combination}, which corresponds to the choice  $K=1$, and
the GMM with $K=2$. The $K=2$ case achieves the lowest overall AIC score at $e_S\approx 1.2$, close
to the nominal $e_S=1$. 

For $e_S=1$, the $K=2$ model returns a weighted
average of two distinct values of $\alpha_s(M_Z)=\left\{{0.1154,
0.1193}\right\}$. As $e_S$ increases, the data become more consistent, thus increasing
the AIC score for $K=2$ and reducing it for
$K=1$. Beyond the value of  $e_S \approx 2.6$, the $K=1$ model would be a 
preferred choice. For  $e_S\approx 2.6$, the fit is still of poor
quality with a large $\sum_i\chi^2(\overline\alpha_s,\alpha_{s,i},\sigma_i)$, hence we do not use this as a
measure to set the value of $e_S$ for $K=1$.  
For values of $e_S \gtrsim 3.5$, the AIC curves for
both models grow linearly with the same slope. For these values of
$e_S$, the $K=2$ model does not find two numerically distinct values
of $\alpha_s(M_Z)$ and therefore behaves like the $K=1$ model,
albeit with some overfitting due to its larger number of parameters.
This also explains a dip in the uncertainty for $K=2$ at $e_S\approx
4.2$ in Fig.~\ref{fig:alpha_s_combination}, above which the
$K=2$ uncertainty starts to track that for $K=1$. The slightly larger
AIC score and uncertainties for $K=2$ for $e_S\gtrsim3.5$ arise due to
overfitting~\cite{Yan:2024yir}. 

\section{Sensitivity to charm-quark mass variations}
\label{app: mc-var}
To investigate the sensitivity of the results to the charm-quark pole mass $m_c$, we extracted $\alpha_s(M_Z)$ from two additional $\chi^2_R$ scans obtained with $m_c=1.2$ GeV and $m_c=1.4$ GeV respectively.  The default charm-quark mass value used in the CT25  global analysis is $m_c=1.3$ GeV.

\begin{figure}
\centering
\includegraphics[width=0.8\linewidth]{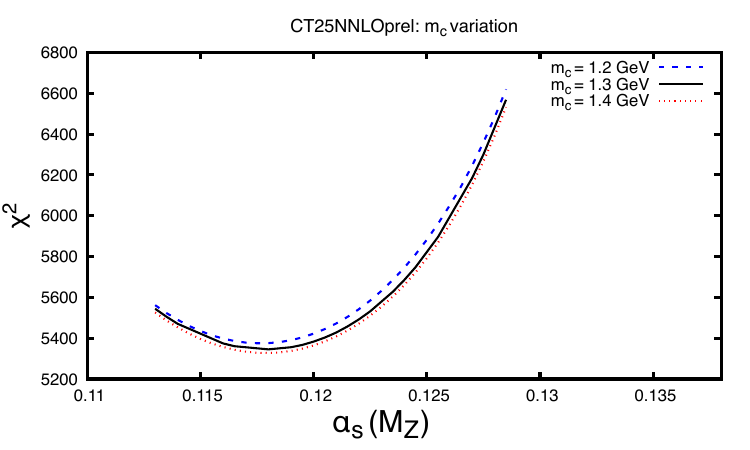}
\caption{$\chi^2$ as a function of $\alpha_s(M_Z)$ in correspondence of the three values of $m_c$ considered in this analysis.}
\label{chi2-vs-alps}
\end{figure}

From the global tolerance criterion, we obtain the following results with the CT25prel baseline:
\begin{eqnarray}
\alpha_s(M_Z) = 0.1175^{+ 0.0025}_{- 0.0023}  \, ,~~~~m_c = 1.2~\textrm{GeV};
\\
\alpha_s(M_Z) = 0.1179^{+ 0.0024}_{- 0.0025} \, ,~~~~m_c = 1.3~\textrm{GeV};
\\
\alpha_s(M_Z) = 0.1179^{+ 0.0024}_{- 0.0025} \, ,~~~~m_c = 1.4~\textrm{GeV}.  
\end{eqnarray}
From this we note that the impact on the uncertainty is negligible, which is also reflected in Figure~\ref{chi2-vs-alps} showing the global $\chi^2$ as a function of $\alpha_s(M_Z)$ for the three values of $m_c$.  

With the dynamical tolerance criterion we obtain at 68\% CL:
\begin{eqnarray}
\alpha_s(M_Z) = 0.1175^{+ 0.0019}_{- 0.0011}  \, , ~~~~m_c = 1.2~\textrm{GeV};
\\
\alpha_s(M_Z) = 0.1179^{+0.0024}_{- 0.0012} \, ,~~~~m_c = 1.3~\textrm{GeV};
\\
\alpha_s(M_Z) = 0.1179^{+0.0025}_{- 0.0013} \, ,~~~~m_c = 1.4~\textrm{GeV}.   
\end{eqnarray}

The DT errors from alternative clustering for the different $m_c$ values behave similarly to those discussed in Sec.~\ref{sec:Tolerances}.


\clearpage

\newpage
\bibliographystyle{apsrev4-2-titles}
%

\end{document}